\documentclass[11pt,preprint2]{aastex}

\shorttitle{Dark \& Baryonic Matter in Galaxies: Distributions}
\shortauthors{Kassin, de Jong, \& Weiner}

\begin{document}

\title{Dark and Baryonic Matter in Bright Spiral Galaxies: II. Radial
  Distributions for 34 Galaxies}

\author{Susan A. Kassin\altaffilmark{1,2},
Roelof S. de Jong\altaffilmark{3}, \&
Benjamin J. Weiner\altaffilmark{4}
}

\altaffiltext{1}{Department of Astronomy, The Ohio State University}
\altaffiltext{2}{currently at: UCO/Lick Observatory, 
University of California, Santa Cruz, CA 95064; kassin@ucolick.org}
\altaffiltext{3}{Space Telescope Science Institute, 3700 San Martin
  Drive, Baltimore, MD 21218}
\altaffiltext{4}{Department of Astronomy, University of Maryland, College Park,
  MD 20742}

\begin{abstract}
We decompose the rotation curves of 34 bright spiral galaxies into baryonic and dark matter components. 
Stellar mass profiles are created by applying color-$M/L$ relations to near-infrared and optical photometry.  
We find that the radial profile of the baryonic-to-dark-matter ratio is self-similar for all galaxies, 
when scaled to the radius where the contribution of the baryonic mass to the rotation curve equals 
that of the dark matter ($R_X$). We argue that this is due to the quasi-exponential nature
of disks and rotation curves that are nearly flat after an initial rise. The radius $R_X$ is found to
correlate most strongly with baryonic rotation speed, such that galaxies with $R_X$ measurements that
lie further out in their disks rotate faster.  This quantity also correlates very strongly with stellar mass, Hubble type,
and observed rotation speed; $B$-band central surface brightness is less related to $R_X$
than these other galaxy properties.  Most of the galaxies in our sample appear to be close to maximal disk.
For these galaxies, we find that maximum observed rotation speeds are tightly correlated with maximum
rotation speeds predicted from the baryon distributions, such that one can create a Tully-Fisher relation based
on surface photometry and redshifts alone.  
Finally, we compare our data to the NFW parameterization for dark matter profiles 
with and without including adiabatic contraction as it is most commonly implemented. 
Fits are generally poor, and all but 2 galaxies 
are better fit if adiabatic contraction is not performed.
In order to have better fits, and especially to accommodate adiabatic contraction,
baryons would need to contribute very little to the total mass
in the inner parts of galaxies, seemingly in contrast with other observational constraints.
\end{abstract}


\section{Introduction}
The flatness of rotation curves has long been the most direct evidence 
for the existence of a dominant component of dark matter in spiral galaxies \citep{rubrev}.  
However, the distribution of dark matter is poorly constrained, 
even in galaxies where the presence of dark matter is dominant
\citep[e.g.,][]{verh,debl}.  
Interestingly, the main uncertainty in the radial 
density distribution of dark matter, as derived from rotation curves, 
stems from the poorly known stellar mass 
distribution \citep[e.g.,][]{verh}. 

In this paper, we do not rely on the typical maximal disk assumption
\citep{vana,vsan} to derive stellar mass profiles, but use color-mass-to-light ratio ($M/L$) relations 
first given by \citet{bdej} and later updated by \citet{bell}.
These color-$M/L$ relations 
were determined from the analysis of spectro-photometric spiral galaxy
evolution models and give an upper limit to the baryonic mass present in
spirals.  They are most powerful when applied to data in the near-infrared
since this wavelength regime traces the older stellar populations that 
contain most of the mass, while avoiding
much of the effects of dust (extinction at the 
near-infrared $K$-band is only about $10 \% $ of 
that at the optical $B$-band; \citealt{mart}).
The color-$M/L$ relations allow us to investigate radial variation
in stellar $M/L$, and provide us with a consistent way of
scaling stellar $M/L$ from one galaxy to the next.
With the distributions of stellar mass in galaxies reasonably well determined,
we can begin to investigate those of the dark matter.

The maximum disk assumption \citep{sack} can be examined with the color-$M/L$ relations.
We can obtain an upper limit to the number of galaxies that have maximal disks,
and also compare the shapes of rotation curves derived from baryonic mass distributions
in galaxies to the observed rotation curves as in \citet{palu}.
There are many pieces of evidence in the literature for maximal
disks; we mention a few noteworthy ones here.  
Possibly the strongest evidence 
comes from the work of \citet{wein} who used fluid-dynamical models
of gas flow to model the two-dimensional velocity field of a barred galaxy.
They find that such models require $80-100\%$ of the maximal disk $M/L_I$ value.
\citet{palu} modeled the baryonic mass distributions of 74 spirals
and found that a mass-follows-light model could reproduce the overall structure
of the optical rotation curves in the majority  of galaxies.  These
authors found that $75\%$ of the galaxies in their sample have a rotation
curve out to $R_{23.5}$ that is entirely accounted for by baryons.
The mass models for $20\%$ of their sample fail because of 
nonaxysymmetric structures.  For a more inclusive discussion on
maximal disks, see \citet{palu}.

It has been known for nearly 20 years that the radius where dark matter
begins to contribute to the rotation curve of a galaxy is 
smaller for low surface brightness galaxies than it 
is for brighter galaxies \citep[e.g.,][]{ps88, ps90, broe, urcp, deb7}.  A number 
of authors have demonstrated this by evaluating either the mass discrepancy
of galaxy disks (ratio of total to baryonic mass) by adopting a stellar $M/L$,
or a similar quantity at a chosen radius.
For example, \citet{urcp} find that spiral galaxies with $V=100$ km s$^{-1}$ 
have $>75\%$ of their mass in dark matter within $R_{opt} \equiv 3.2 h$ where $h$ is the disk
scale-length, and that galaxies
with $V=150$ km/s have only $>40\%$.  \citet{mc98} find that the total
$M/L_B$ evaluated at $4h$ is smaller for galaxies of higher surface brightness
and brighter magnitude, and does not correlate with $h$.  These authors also find that
the radius where dark matter begins to dominate over the luminous mass
is greater for higher surface brightness galaxies.  \citet{zava}
find that the ratio of the velocity due to the baryons to the total maximum velocity 
depends mainly on disk central surface density such that
denser galaxies have larger ratios.  
These authors also find that the mass discrepancy evaluated at $2.2h$ and $5h$
depends on central surface brightness,
but puzzlingly does not depend on $h$, baryonic mass, or $B$-band luminosity.  In addition, \citet{zava} find that $M/L_B$ 
anti-correlates significantly with $B$-band central surface brightness and does not
correlate with $B$-band luminosity, in disagreement with \citet{mc98} and the general
qualitative trend in the literature.
Most recently, \citet{piza} evaluate the total $M/L_i$ at $2.2h_i$
and find median values of 2.4 and 4.4 for galaxies with stellar masses greater than
$ 10^{10} M_{\odot}$ and 4.4 between $10^9$--$10^{10} M_{\odot}$, respectively.
These authors are all generally in qualitative agreement, but there are 
discrepancies, and we will discuss their origin in the body of this paper.

A combination of resolved H$\alpha$ or CO and \ion{H}{1} velocity data is 
necessary to determine the dark matter distribution of a galaxy from its
center to beyond $\sim 5h$.  H$\alpha$ data are essential to measure the steep 
rise of rotation curves in the inner $\sim2h$ of galaxies \citep[e.g.,][]{palu},
whereas \ion{H}{1} data are key to probe
beyond where H$\alpha$ can be measured \citep[e.g.,][]{deb7}.  Since rotation curves in the
optical portion of galaxies are usually affected by non-axisymmetric
structures that do not trace the main galactic potential,
such as bars and spiral arms, the distributions of dark matter derived
from H$\alpha$ data can be quite variable.  Two examples of studies probing dark matter
distributions that have used both types of rotation curves in a direct manner, 
independent of prior parameterizations for the dark matter,
are \citet{urcp} and \citet{mc98}. \citet{urcp} coadd 1100 rotation curves 
and find that they could be determined by a single galactic parameter (e.g., luminosity).  This
formulation, however, is refuted by a number of authors \citep[e.g.,][]{verh,bosm}.
 \citet{mc98} find a regularity between the mass
discrepancy and acceleration of a galaxy as a function of radius.
This follows directly from the Tully-Fisher relation, probably the best observational 
example of self-similarity in the dark matter component of galaxies.
The mass discrepancy-acceleration relation shows that knowledge of the baryonic mass distribution
of a galaxy allows one to calculate its dark matter distribution.
Such a relation steps beyond the usual Tully-Fisher relation, which
predicts the dark matter content of a galaxy at a particular
radius given its baryonic mass at that radius, to one that predicts 
radial distributions.

Theories and simulations of galaxy formation should be able 
to reproduce the distributions of dark matter observed in galaxies.
Until now, comparisons to predictions for dark matter distributions have consisted
of fits of functional formulations for dark matter halos to rotation curve data, with 
stellar $M/L$ as a free parameter.  Now that we have a handle on the stellar
$M/L$ in galaxies with the color-$M/L$ relations, we re-examine these fits.
In particular, we test the main incarnation of the density distribution of dark
matter halos in N-body simulations of a $\Lambda$CDM universe: the formulation of
\citet{nnfw} (NFW), which is the simplest and most popular analytical description.
This formulation predicts that dark matter is significant 
down to the very inner radii of galaxies, in contradiction to
maximal disks.  Moreover, dark matter halos are expected to contract due to baryon collapse.
As this contraction is normally implemented \citep{blum},
even more dark matter is caused to move toward the centers of galaxies, making the 
situation worse \citep[e.g.,][]{mc98}.  We perform fits to the NFW formulation with and without 
including contraction as it is normally implemented.  The main assumption behind these fits, namely that
the NFW halos can be fit to data that subtends only their very inner parts 
($\lesssim 0.1 R_{virial}$), is problematic.
Furthermore, the simulations that produce the NFW functions have difficulty forming
disk galaxies once hydrodynamics is included \citep[e.g.,][]{abad},
signaling possible unknown interactions between baryons and dark matter.
To address these concerns, we turn the problem of fitting dark matter 
halos to data around and present a simple universal form for dark matter profiles
in terms of baryonic mass profiles.

This paper is organized as follows: In $\S$2 we briefly discuss 
our galaxy sample.  Radial baryonic mass distributions and their 
rotation curves are derived in $\S$3.  In $\S$4 these baryonic rotation
curves are compared with rotation curves from
the literature to derive rotation curves due to dark matter.  
We discuss maximal disks in $\S$5.
In $\S$6 we define quantities to describe 
the baryonic and dark matter distributions and show how they 
correlate with general galaxy properties.  
We investigate the radial behavior of dark matter in $\S$7, and
mass profiles are fit with the NFW profile in $\S$8. 
In $\S$9, we summarize our conclusions.
Throughout this paper we adopt a Hubble constant ($H_o$) of 70 
km s$^{-1}$ Mpc$^{-1}$. When distance-dependent quantities 
have been derived from the literature, we
have reduced them to this value of $H_o$ and always quote the
converted values.

\section{The Galaxy Sample}
Our data consist of surface brightness profiles, physical
parameters, and rotation curves for 34 bright spiral galaxies.
These galaxies have inclinations in the range
$\sim 30$--65 degrees in order to
reduce the effects of dust, while still being able to obtain accurate
kinematical information.  
Surface brightness profiles for 30 galaxies were presented in \citet{kass} 
(hereafter, Paper\,I); those for the remaining 4 galaxies can be found in \citet{verh}.
More details on the sample selection are given in Paper\,I.
The profiles were calculated 
in elliptical annuli of increasing distance from the centers of the
galaxies and have been corrected for Galactic extinction.
The galaxies NGC\,1090, NGC\,2841, 
and NGC\,3198, which have images from the Sloan Digital Sky Survey (SDSS)
Second Data Release \citep{abaz}, have images in which almost half of the 
galaxy was off of the detector.  This is also the case, but to a much 
lesser extent for NGC\,3521.  This effect can be observed
in the $g$-band images in Figure 1 of Paper\,I; these galaxies are
flagged in the following analysis.  All the analyses
in this paper have been performed for both the approaching
and receding sides of a handful of galaxies
with complete imaging.  No significant change (to within the zero-point uncertainties)
was found between the results for each side of these galaxies. 
Physical parameters (i.e., Hubble type, distance,
$R_{25}$, integrated magnitudes) and bulge-disk decompositions
can also be found in Paper\,I.

\section{Baryonic Matter}
\label{sec:baryonicmatter}
In this section, we derive the
baryonic mass surface density profiles of the galaxies, and compute the
component of the galaxies' observed rotation that is due to the
baryons.  These components are called 
``baryonic rotation curves,'' to distinguish them from observed 
rotation curves.
In $\S$4, we will introduce the term ``dark matter rotation curves,''
to denote the contribution from dark matter to the observed
rotation curves. 

\subsection{Radial Baryonic Surface Mass-Density Distributions}
\begin{figure}[center]
\figurenum{1}
\includegraphics[height=2.5in, width=2.9in]{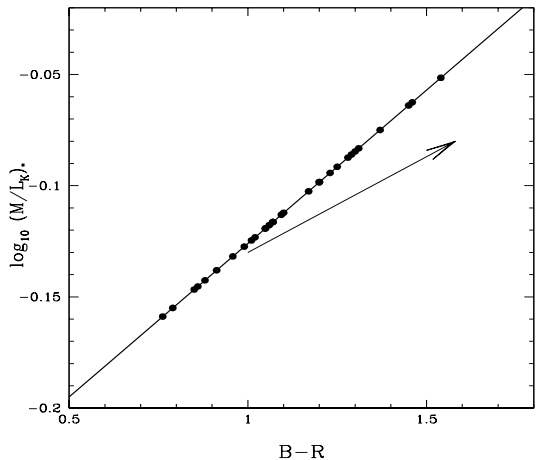}
\caption{{\small Color-$M/L$ relation for $B-R$ color and $(M/L)_*$ at
  $K$ from \citet{bell} (solid line).  The galaxies in the sample are
  plotted as filled circles, and a reddening vector is plotted for
for correction to face-on for a Milky Way-type galaxy viewed at an 
inclination of $80\degr$ with the Tully et al.\ 1998 formalism.}
\label{fig:BdeJ}}
\end{figure}

To determine a galaxy's radial stellar surface mass-density distribution,
we apply a color-$M/L$ relation to its surface
brightness profiles.  These relations were derived from spectrophotometric
spiral galaxy evolution models in \citet{bdej}, and were updated in \citet{bell}.
The color-$M/L$ relations show a relatively tight correlation
($\sim0.1$ dex spread for the color and metallicity range where our galaxies lie)
between the optical color of a galaxy and its stellar mass-to-light
ratio, $(M/L)_*$.  Color-$M/L$ relations are most useful when applied to
$(M/L)_*$ in the near-infrared since they are the least
affected by dust obscuration, as discussed in \citet{bdej}.
Specifically, we choose to use the relation between $B-R$ color
and $(M/L)_*$ in the $K$-band.  This relation is composed of the optical color
with the largest wavelength baseline and the reddest near-infrared band.
It is reproduced in Figure~\ref{fig:BdeJ}.  If imaging is unavailable at $R$,
we use the relation for $B-V$, and if imaging is unavailable at $K$ we use $H$.

For those galaxies without a significant bulge contribution (such that
inclusion of a bulge component does not change the
rotation curve due to baryons beyond the uncertainties of the color-$M/L$ relations), 
we apply a color-$M/L$ relation directly to the azimuthally averaged
radial $B-R$ color profiles to derive radial $(M/L)_*$ profiles at
$K$.  The $L_K$ profile for each galaxy is
then multiplied by the galaxy's $(M/L)_*$ profile to derive a radial stellar surface
mass-density profile.  For these galaxies, using a radial $B-R$
color profile is consistent with using an aperture $B-R$ color
since this color-$M/L$ relation has a shallow slope. 
For those galaxies with a significant bulge component,
a bulge-disk decomposition is performed.  A characteristic $B-R$ color
is adopted for each component based on the average colors of the bulge
and disk, and the $L_K$ profiles are multiplied by the resulting $(M/L)_*$.  
Extinction corrections are discussed in $\S$3.4.  We also extend the 
exponential disk surface brightness profiles until approximately 
$\infty$ (defined here as $r=10000\arcsec$).\footnote{Extended
galactic stellar disks have been discovered for some galaxies via very deep imaging
\citep[see][and references within]{ibat}.}  

\begin{figure}[center]
\figurenum{2}
\includegraphics[height=2.5in, width=2.9in]{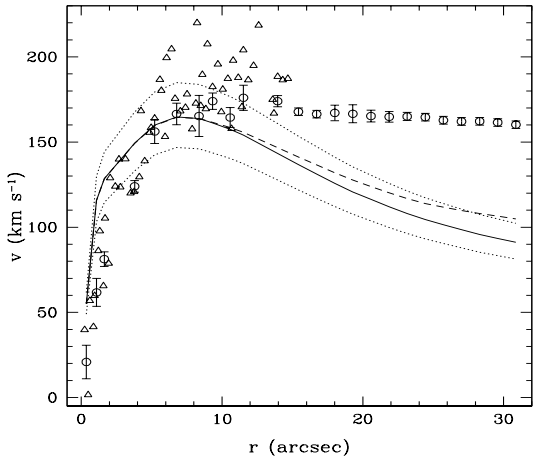}
\caption{{\small Rotation curve for NGC\,1090 due to the 
stellar mass component (solid line) is compared with the 
rotation curve due to stars and \ion{H}{1} gas from Gentile et al. 2004
(dashed line).   The observed H$\alpha$ and \ion{H}{1} 
rotation curves are plotted as open triangles and circles,
respectively.   The upper and lower bounds for
the stellar mass rotation curve (dotted lines) are due to 
the 0.1 dex uncertainty in the color-$M/L$ relations.}\label{fig:Gas_Added}}
\end{figure}

When available from the literature, we use radial \ion{H}{1}~21\,cm
measurements to determine the contribution of interstellar gas
to the radial baryonic surface mass-densities.  There are gas measurements available
for NGC\,1090 from \citet{gent}, NGC\,3198 from \citet{bege}, and 
NGC\,3949, NGC\,3953, and NGC\,3992 from \citet{verh}.
The neutral gas component is included by
scaling the \ion{H}{1} surface mass-density by a factor of 1.32 to account for the
abundance of helium.  For those galaxies without gas mass measurements,
we assume that gas does not contribute to its baryonic surface mass-density.
This is probably not a bad assumption for high surface brightness galaxies since
a galaxy with a Hubble type between Sa and Scd 
has a gas mass fraction that is typically $M_{HI}/M_{gas + stars} \approx 0.03$ \citep{robe},
and hence does not greatly affect its baryonic
mass.  The average gas mass fraction for the 5 galaxies in our sample
that have gas mass measurements is 0.04, consistent with this measurement.
Figure~\ref{fig:Gas_Added} demonstrates the effect of
including the interstellar gas component in the baryonic 
rotation curve of NGC\,1090, which is generally within the 
uncertainty of the stellar mass rotation curve.  Therefore, by
not including interstellar medium contributions in the analysis,
only a minor systematic effect is introduced.  
\subsection{Baryonic Rotation Curves}
A baryonic rotation curve is calculated for each galaxy
from its baryonic surface mass-density profile and
is plotted in Figure 3. 
The rotation curve calculation is done with the $\tt rotmod$ task in the
GIPSY software package \citep{gips}, which calculates
rotation curves for galaxies composed of a truncated
exponential disk \citep{case} and a spherical bulge when applicable.  Exponential disks are
assumed to have a scale-height of 0.3 kpc,
which is typical of bright spirals.  The difference between using a
different scale-height for each galaxy of $0.1 h_{IR}$ kpc,
as in \citet{spar}, and a global value of 0.3 kpc 
is negligible.  The results also remain unchanged if we used the 
relation between the central surface brightness and the ratio of 
vertical scale-height to vertical scale-length given by \citet{bizy}.
\begin{figure*}
\figurenum{3a}
\includegraphics[scale=0.8]{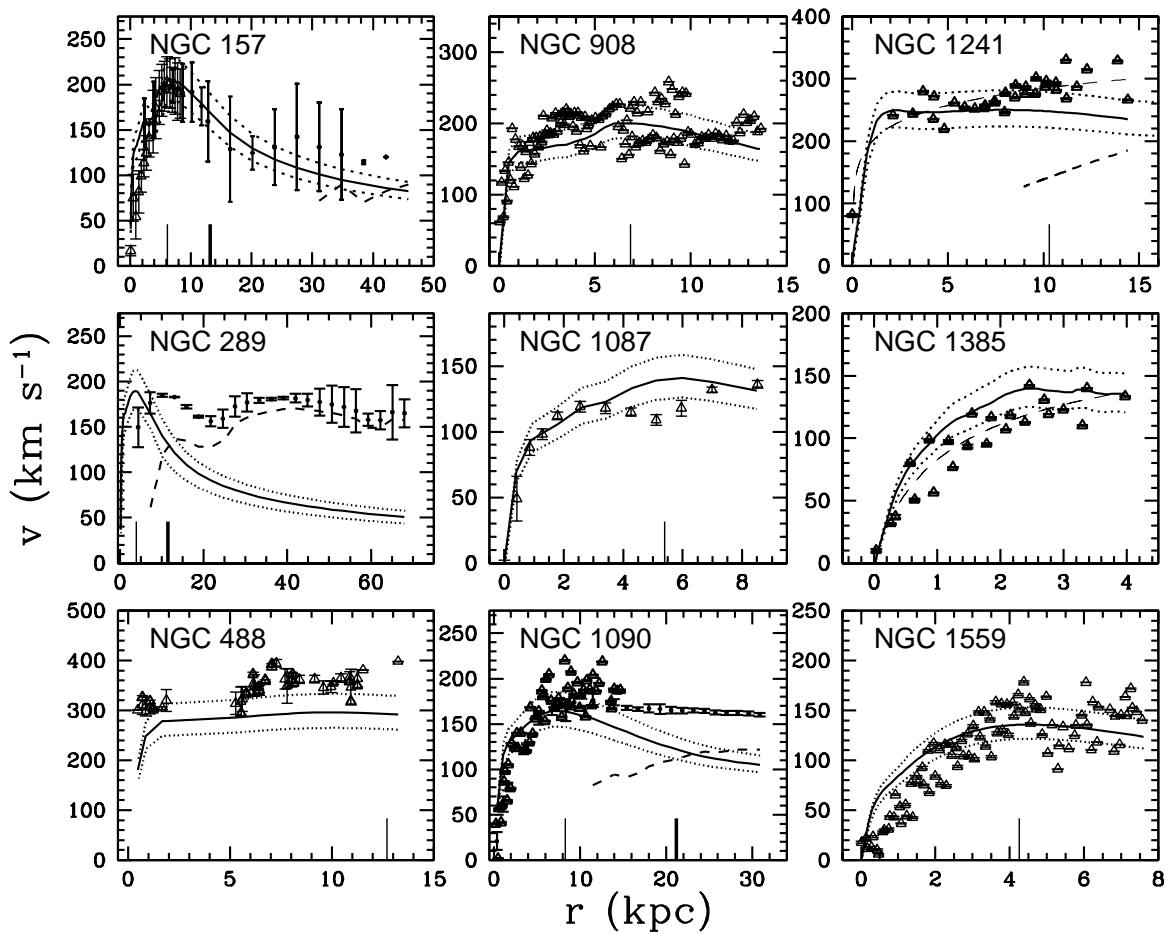}
\caption{{\small Observed rotation curves (dots for \ion{H}{1}, triangles for
  H$\alpha$ or \ion{N}{2}, thin dashed lines for models), baryonic rotation curves
(solid lines), and dark matter rotation curves where applicable
  (thick dashed lines).  
For many of the H$\alpha$ and \ion{N}{2} rotation curves,
the error bars are smaller than the points at the resolution of the plots.
The effects of the $\pm 0.1$ dex uncertainty in the color-$M/L$ relations on the
baryonic rotation curves are plotted as dotted lines.
On the x-axis, the radii $R_{25}$ and 
$R=2.2 h_{IR}$ are marked with thick and thin bars, respectively.}
\label{fig:rc}}
\end{figure*}
\begin{figure*}
\figurenum{3b}
\includegraphics[scale=0.9]{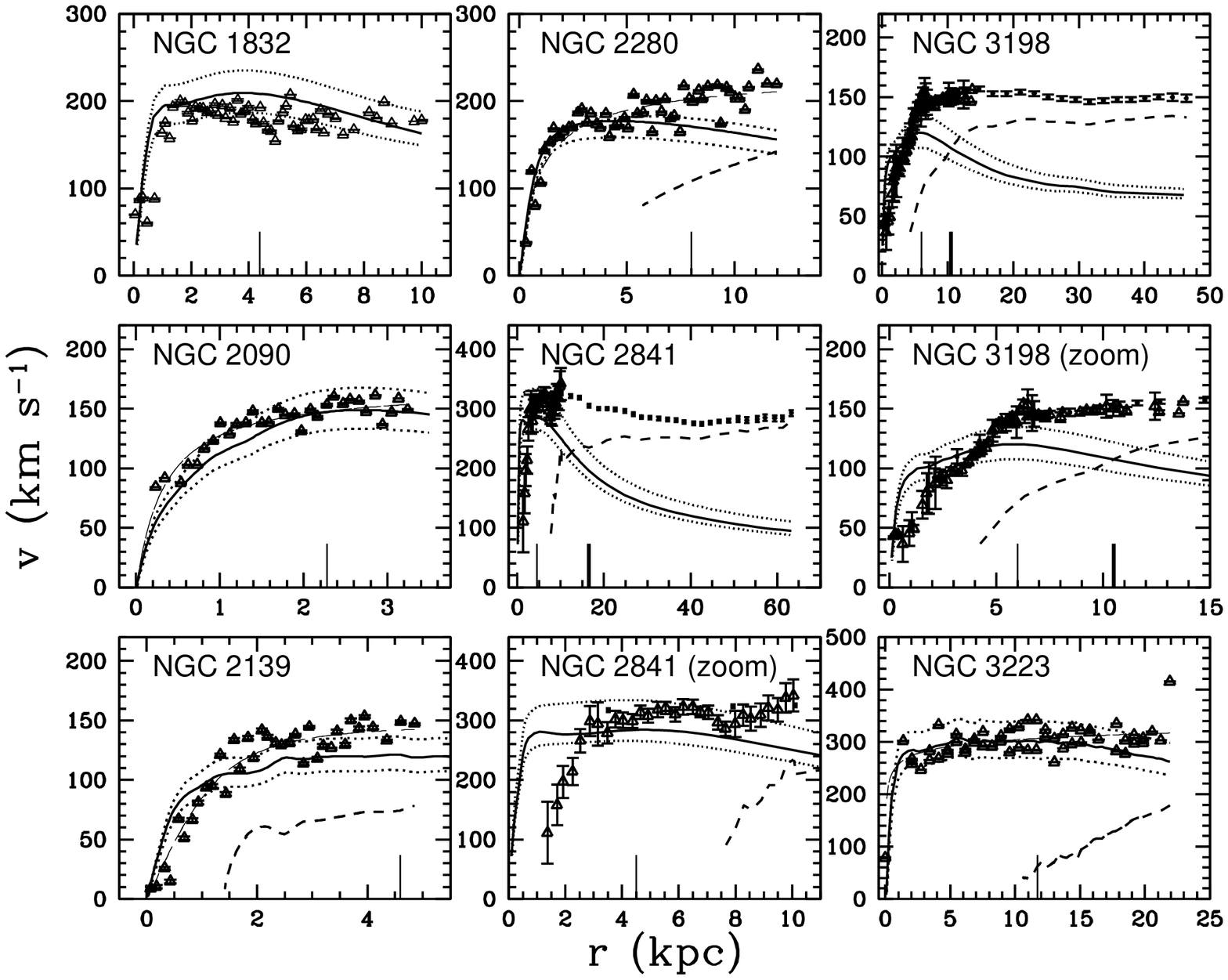}
\caption{{\small See Figure 3a.}}
\end{figure*}
\begin{figure*}
\figurenum{3c}
\includegraphics[scale=0.9]{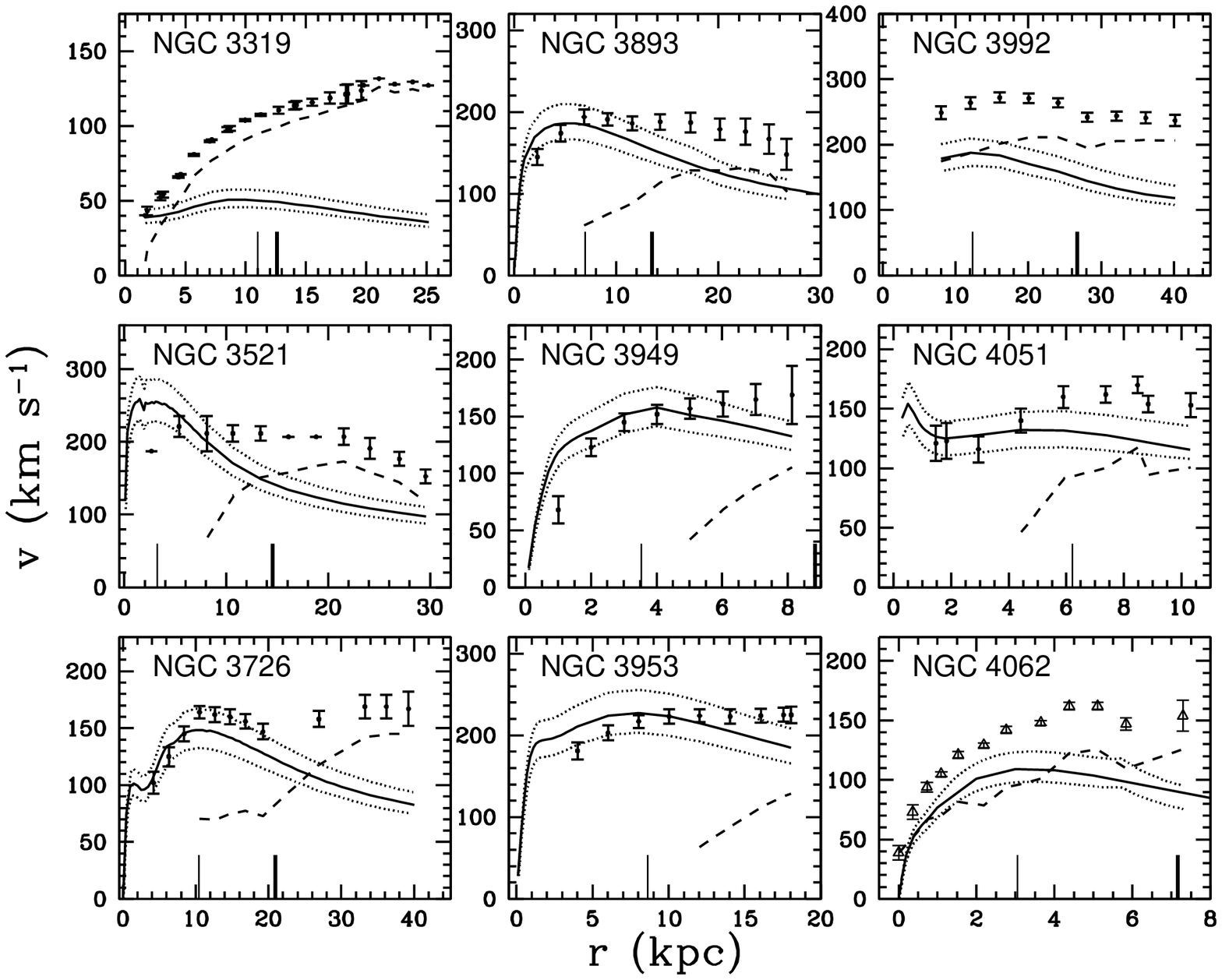}
\caption{{\small See Figure 3a.}}
\end{figure*}
\begin{figure*}
\figurenum{3d}
\includegraphics[scale=0.9]{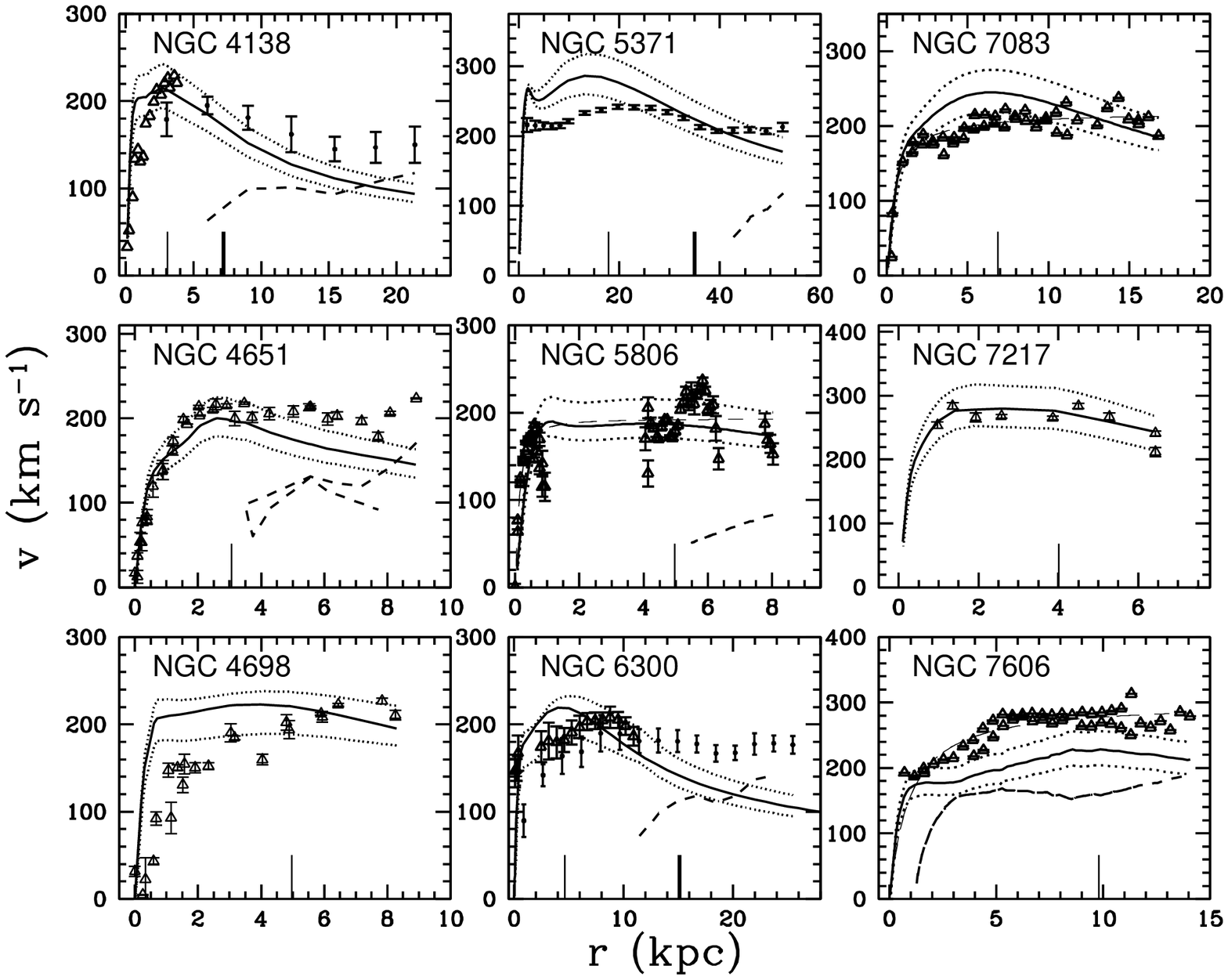}
\caption{{\small See Figure 3a.}}
\end{figure*}
\subsection{Uncertainties in Baryonic Rotation Curves}
\label{sec:uncbaryon}
The largest source of uncertainty for the baryonic rotation curves lies in
the normalization of the color-$M/L$ relation, which is mainly determined by
the stellar IMF at low-mass.  Since the faint end of the
IMF is relatively unconstrained, there may exist many low-mass,
low-luminosity stars that can contribute significantly to the mass
budget of a stellar population without creating a detectable
increase in luminosity or change in color. 
For their derivation of the color-$M/L$ relations,
\citet{bdej} and \citet{bell} adopted a truncated Salpeter IMF which derives from the
constraint that baryonic rotation curves should not over-predict
observed rotation curves for spiral galaxies in Ursa
Major (the Verheijen 1997 sample).  With this constraint, they predict
fewer low mass stars than a Salpeter IMF.
These relations thus give an {\it upper limit} to the stellar mass
present; a lower normalization of the color-$M/L$ relation cannot be excluded.

\begin{figure}[center]
\figurenum{4}
\includegraphics[height=2.9in, width=2.9in]{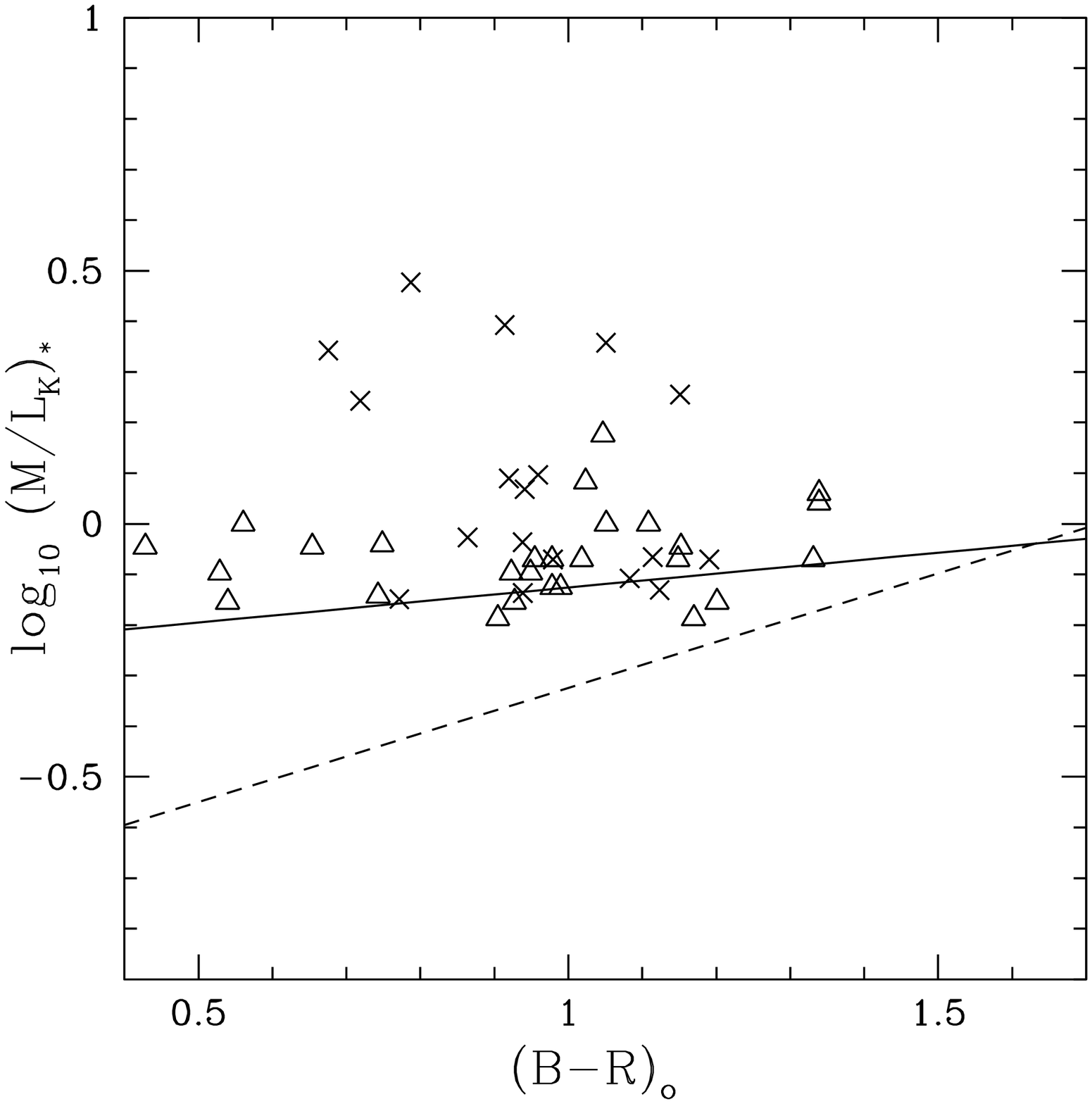}
\caption{{\small Maximum disk stellar $(M/L_K)_*$ versus reddening-corrected $B-R$ colors for galaxies in the \citet{verh}
sample (Xs) and in this paper (triangles).
Plotted for reference are the 
color-$M/L$ relations of Bell \& de Jong 2001 (dashed line) and 
Bell et al.\ 2003 (solid line).}
\label{fig:BdeJ_upperlim}}
\end{figure}

We re-derive this constraint on the upper limit to the IMF 
by re-calculating maximum disk fits for galaxies in the \citet{verh} sample without including
a dark halo, as was done in the original \citet{verh} fits.  
For these fits, we define maximal disk to be the greatest possible contribution
of the bulge and disk to the observed rotation curve of the galaxy.
To do this, we scale the $K$-band derived stellar mass rotation curves as 
high as possible without over-predicting the observed rotation curves beyond 
the very inner parts.  In order to best match the overall shapes of the observed rotation
curves, we allow over-predictions of the inner 2 points of 
\ion{H}{1} rotation curves due to beam-smearing and over-predictions of the inner 
parts of H$\alpha$ rotation curves.    We also re-scale the 
stellar and gas mass distributions from the distance used in \citet{verh} 
to 20.7 Mpc before fitting
the maximal disks, unlike the approximate re-scaling done in \citet{bdej}.
Gas distributions derived from \ion{H}{1} measurements are multiplied
by 1.32 to account for helium, as in \citet{verh}.  In Figure~\ref{fig:BdeJ_upperlim},
the resulting maximum disk $(M/L)_*$ are plotted against the 
$B-R$ colors of the galaxies corrected for extinction with the \citet{tull} formalism.  In 
this figure, we also plot maximum disk $(M/L)_*$ for those galaxies presented in
this paper that have $BRK$ photometry.  All the points in Figure~\ref{fig:BdeJ_upperlim} are upper limits
to the $(M/L)_*$; galaxies cannot have values greater than those defined by the
lower envelope in this plot without over-predicting their rotation curves beyond the very inner parts.   
It is apparent from Figure~\ref{fig:BdeJ_upperlim} that the 
\citet{bell} color-$M/L$ relation is consistent with all the galaxies to within $\sim0.1$ dex, 
which is the uncertainty in the relations.  

Other than the IMF, there are five primary sources of uncertainty in the determination of
the baryonic surface mass-density distributions: uncertainties
in the distances to galaxies\footnote{Of the 34 galaxies in our sample, 4
have distances estimated from Cepheid variables for which the
uncertainties are much less than other types of distance measurements, and
are typically $\sim 10$\%.}, 0.1 dex spread in the color-$M/L$ 
relations, uncertainties in the zero-points of these relations,
uncertainties in the photometric zero-point calibrations,
and uncertainties in the determination of a galaxy's inclination to the
line of sight.  Secondary sources of uncertainty generally have a small effect on the 
baryonic rotation curves.  They include secondary photometric
(``bootstrap'') calibrations, the neglect of interstellar gas
content, and position angle uncertainties.  Dust reddening
also plays a role in the uncertainty of the baryonic
rotation curves; this will be discussed in $\S$4.3.
Note that the transformation of Two Micron All Sky Survey 
\citep[2MASS;][]{jar2,cutr,jar1} and SDSS photometry
to the Kron-Cousins system introduces negligible
uncertainties, as shown in Paper\,I.

\begin{figure}[!t]
\figurenum{5}
\hspace{-0.1in}
\includegraphics[height=3.in, width=3.3in]{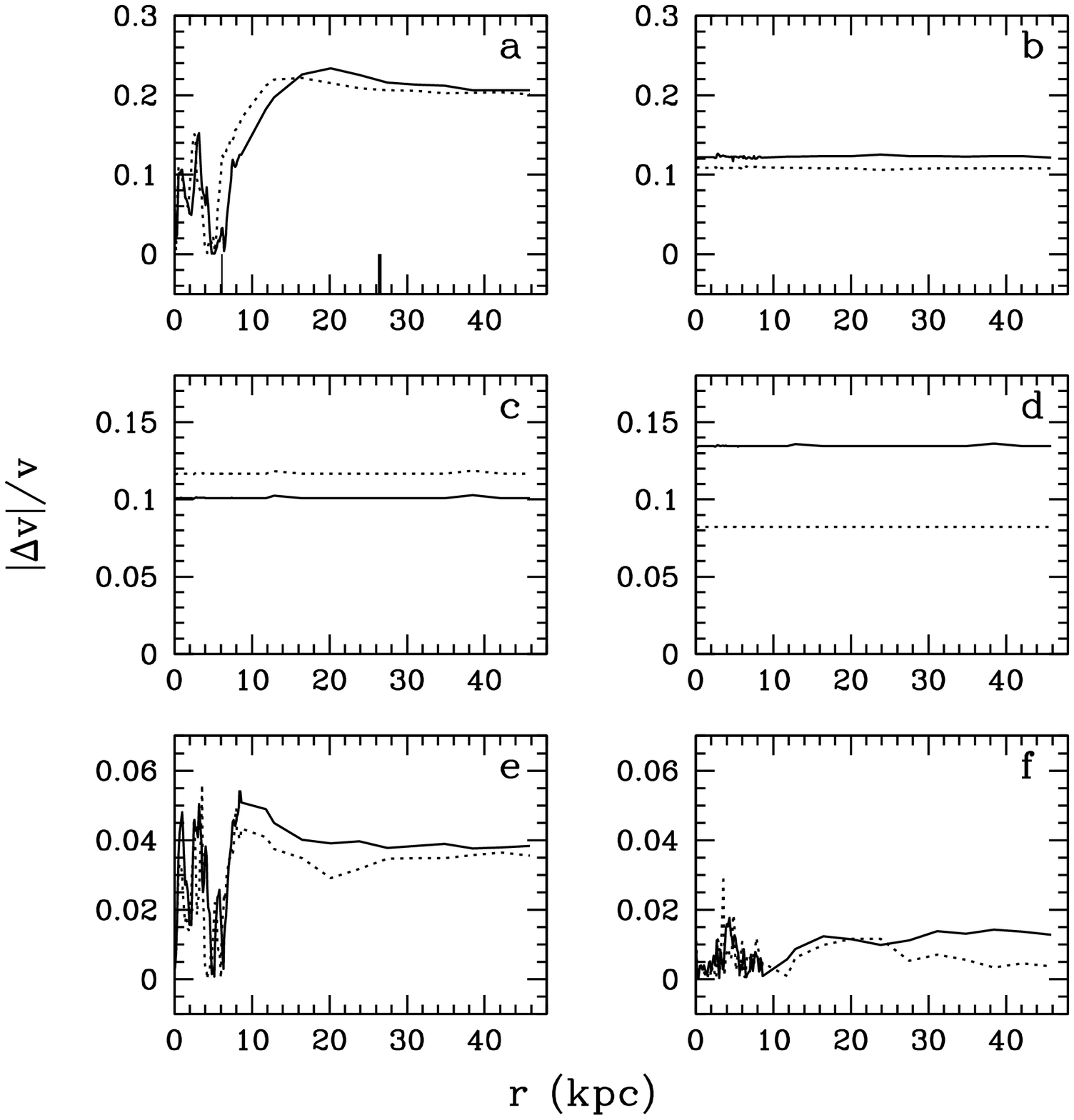}
\vspace{-0.3in}
\caption{{\small Effects of uncertainties on the
baryonic rotation curve of NGC\,157.  
For one side of the uncertainty (e.g., +0.1 dex), the $|\Delta v_b| / v_b$ curve is
plotted as a dotted line, and for the other side (e.g., -0.1 dex) the curve
is plotted as a solid line.  The effects of the following uncertainties
are plotted: (a) $\pm 20\%$ uncertainty in the distance to the galaxy,
(b) $\pm 0.1$ dex scatter in the color-$M/L$
relation,  (c) actual photometric zero-point uncertainties for NGC\,157,  (d) 
photometric zero-point uncertainties for the galaxy with the poorest zero-point,
(e) a change in the inclination of $5\degr$, and 
(f) a change in the position angle of $5\degr$.
The radii $2.2 h_K$ and $R_{25}$ are marked as thin and thick lines,
respectively, on the x-axis in panel a.}
\label{fig:Uncertainties}}
\end{figure}

As an illustration, Figure~\ref{fig:Uncertainties}a--d
shows the effects of three of the primary
sources of uncertainty on the baryonic rotation curve of NGC\,157.
In all the plots in Figure~\ref{fig:Uncertainties}, we plot $|\Delta v_b| / v_b$
where $\Delta v_b$ is the difference between the velocity derived from the
color-$M/L$ relations and this velocity affected by the named uncertainty.
Figure~\ref{fig:Uncertainties}a shows the effect of an uncertainty of $\pm \ 20\%$
in the distance.  The distance uncertainty causes a maximum change in $|\Delta v_b| / v_b$ of 0.24 which
corresponds to a change in the velocity of
nearly 40 km s$^{-1}$.  The average change is $|\Delta v_b| / v_b=0.09$ 
($|\Delta v_b| \backsimeq 13$ km s$^{-1}$).
Figure~\ref{fig:Uncertainties}b shows the effect of a systematic
change of $\pm 0.1$ dex in the color-$M/L$ relations.  
This uncertainty causes a maximum change in $|\Delta v_b| / v_b$ of 0.12
($|\Delta v_b| \backsimeq 25$ km s$^{-1}$), and an average change of 0.12 
($|\Delta v_b| \backsimeq 20$ km s$^{-1}$).  
In Figure~\ref{fig:Uncertainties}c,d, the effects of introducing systematic
errors in the photometric zero-point calibrations are shown.  
In Figure~\ref{fig:Uncertainties}c, the effect of the actual
zero-point uncertainties for NGC\,157 is shown ($\sigma_B=0.03, \sigma_R=0.03,
\sigma_K=0.04$).  This causes an average 
change in $|\Delta v_b| / v_b$ of 0.10 ($|\Delta v_b| \backsimeq 17$ km s$^{-1}$).
While most of our optical photometric zero-points are good to $\le \rm5 \%$,
the worst uncertainty in an optical zero-point calibration for any of the
surface brightness profiles that we use in this analysis
is $\pm 15 \%$, as given in Table 3 of Paper\,I.  All of the uncertainties
on the near-infrared zero-points are $\sim 4\%$.
To show the effect of a photometric calibration that is not as good
as that of NGC\,157, in Figure~\ref{fig:Uncertainties}d we show what would happen to
the rotation curve if photometric zero-point
errors were $\pm 15\%$ in the optical and $\pm 4\%$ in the near-infrared.  
Such uncertainties cause an average change in $|\Delta v_b| / v_b$ of 0.13
($|\Delta v_b| \backsimeq 22$ km s$^{-1}$).

In Figure~\ref{fig:Uncertainties}e,f, we show the effects 
on the baryonic rotation curve of NGC\,157 of changing
the inclination and position angle in the derivation of its
surface brightness profiles.
Typical uncertainties in position angles and inclinations are both 
$\sim \pm 5\degr$ (Paper\,I).
In Figure~\ref{fig:Uncertainties}e, we plot 
the effect of a change in the position angle of $\pm 5\degr$, and
in Figure~\ref{fig:Uncertainties}f we show
the effect of a change in the inclination of $\pm 5\degr$.  Uncertainties in the 
inclination have a larger effect (on average $|\Delta v_b| \backsimeq 0.03$; $|\Delta v_b| \backsimeq 5$ km s$^{-1}$) than 
those in the position angle (on average $|\Delta v_b| \backsimeq 0.007$; $|\Delta v_b| \backsimeq 1$ km s$^{-1}$).
However, both these uncertainties are small when compared to other sources of error
discussed in this section.

\begin{figure}[center]
\figurenum{6}
\hspace{-0.2in}
\includegraphics[height=3.5in, width=3.3in]{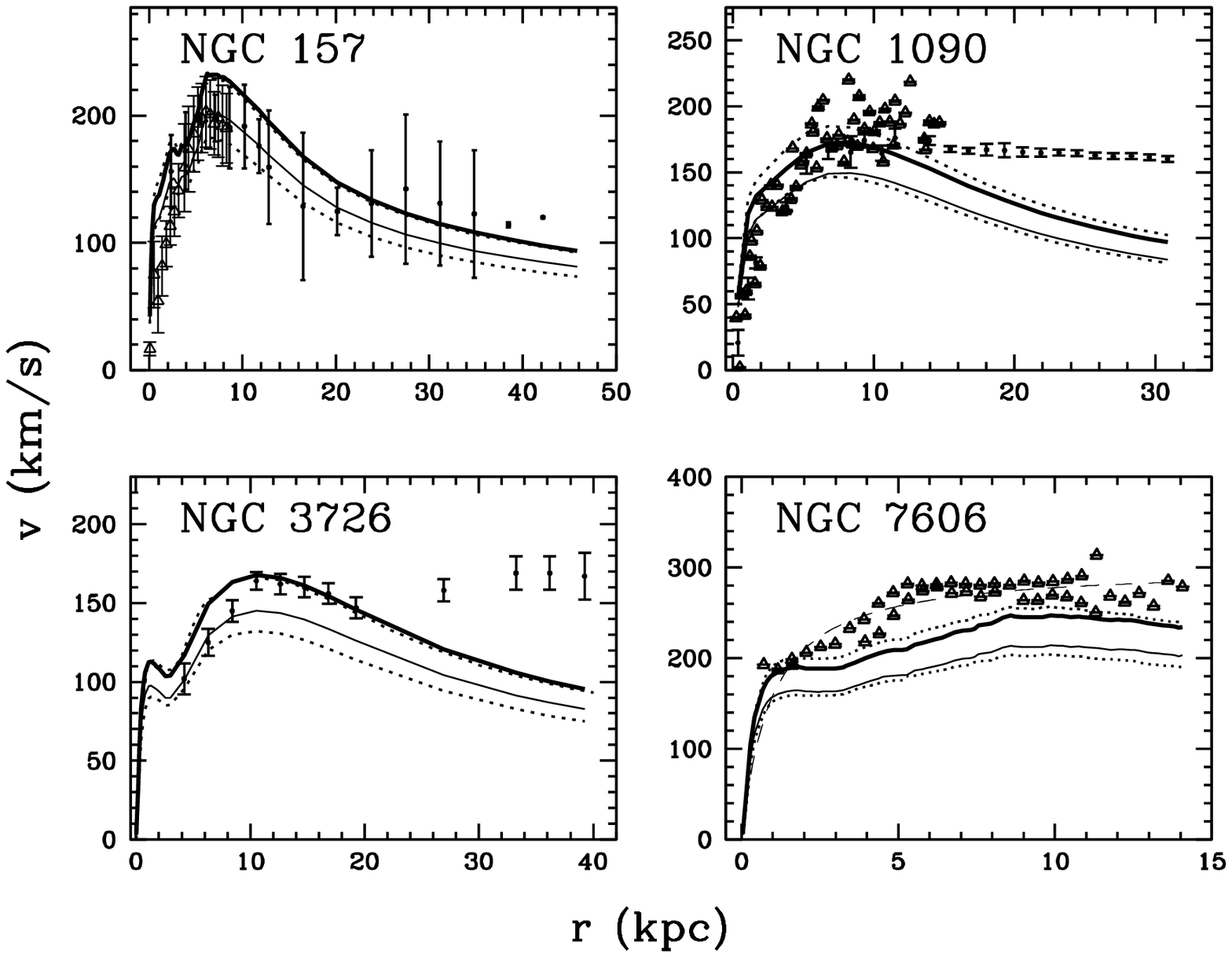}
\vspace{-1.1in}
\caption{{\small Baryonic rotation curves derived from the color-$M/L$ relations 
(plotted as dotted lines for the $\pm 0.1$ dex scatter in the relations)
are compared with baryonic rotation curves derived from constant $(M/L_K)_*$
of 0.75 (thin solid line) and 1.0 (thick solid line).  The thin dashed line
plotted for NGC\,7606 is the model observed rotation curve of Courteau 1997.  All other features of
the plot are the same as Figure 3.}\label{fig:rc_ML}}
\end{figure}

In Figure~\ref{fig:rc_ML}, to compare the color-$M/L$ relations to 
constant $(M/L_K)_*$, baryonic rotation curves created with the color-$M/L$ relations
and with $(M/L_K)_*$ of 0.75 and 1.0 are plotted for 4 example galaxies.  
The rotation curves created with $(M/L_K)_*$ of 0.75 and 1.0 
are approximately consistent with the $\pm0.1$ dex uncertainty
of the color-$M/L$ relations.
This is similarly the case for all galaxies in our sample, and
is due to the shallow slopes of the color-$M/L$ relations 
($\sim 0.14$--$0.18$ in log$_{10} M/L$).

To summarize, other than the IMF, of the other sources of uncertainty in the determination
of the baryonic rotation curves, only the distance
uncertainty, the $\pm0.1$ dex scatter in the color-$M/L$ relations, and zero-point
uncertainties can produce non-negligible effects.
These three sources of uncertainty vary from galaxy to galaxy and introduce scatter in
the baryonic rotation curves.

\subsection{Effects Due to Dust}
Since the reddening vector in Figure~\ref{fig:BdeJ}
lies nearly parallel to the color-$M/L$ relation, to first order,
errors in foreground dust reddening estimates should not strongly
affect the final relative derived masses of the stellar populations, as
foreground dust will systematically both redden and extinguish galactic light.
In this section, we discuss the possible effects of dust reddening and
extinction on absolute derived stellar masses.

\begin{figure}[center]
\figurenum{7}
\hspace{-0.2in}
\includegraphics[height=3in, width=3.3in]{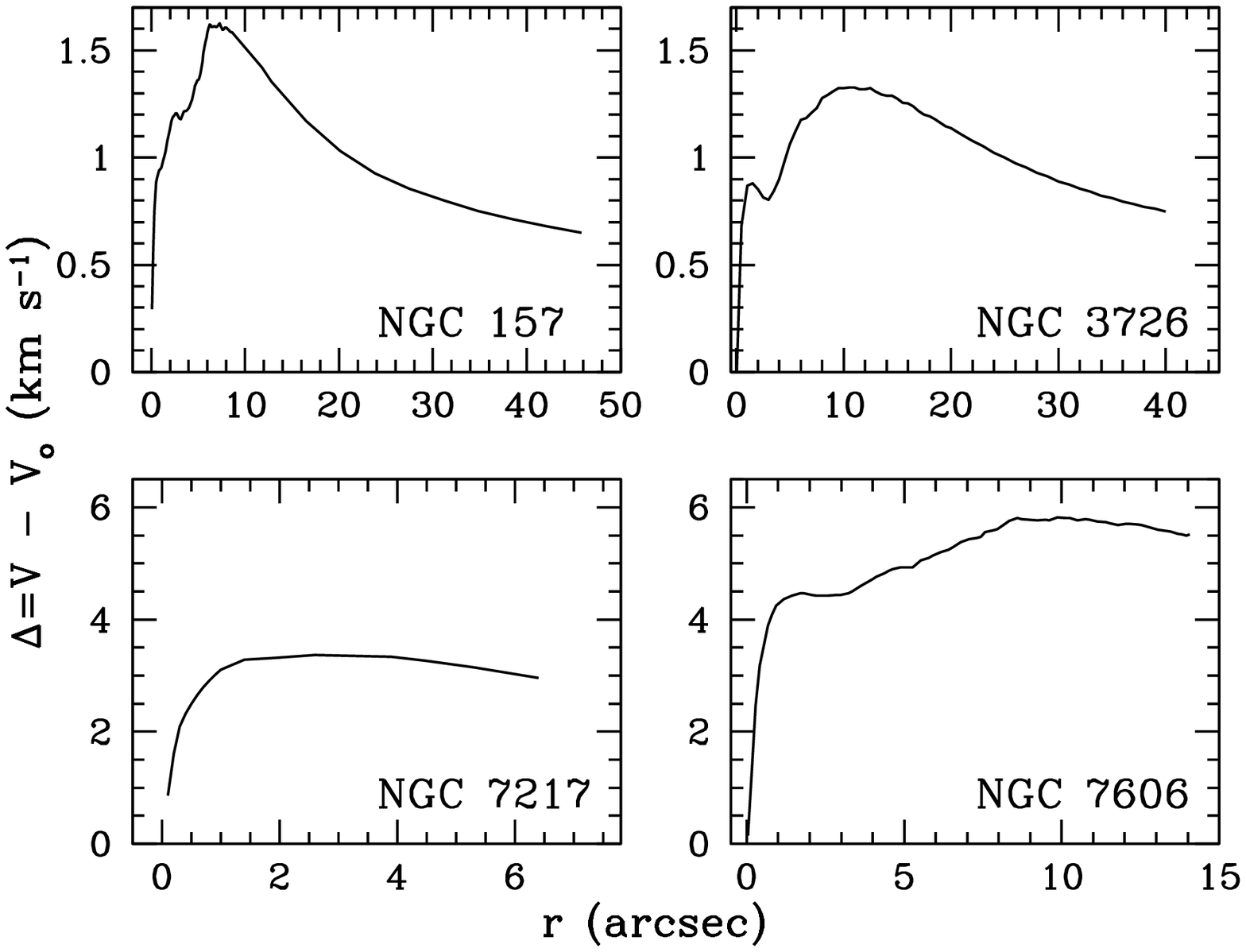}
\vspace{-1.0in}
\caption{{\small Difference between the extinction-corrected ($V$) 
and uncorrected ($V_o$) baryonic rotation curves. The extinction-corrected
baryonic rotation curves are derived from
colors and magnitudes corrected for extinction with
the formalism of Tully et al.\ 1998.} \label{fig:T98_rcs}}
\end{figure}

We examine the empirical inclination-dependent extinction correction of \citet{tull}, 
which should describe to first order the dust content of galaxies.
We correct to face-on the total integrated colors 
and magnitudes of galaxies in our sample.
In doing this, we ignore radial dust gradients
in the disk and the fact that dust reddening is likely to be large for the inner parts of 
galaxies.  While the $K$-band magnitudes barely
change when extinction corrected (0.05 mag on average), the $B-R$ colors do (0.09 mag on average).  
This, however, should not affect the
final derived stellar masses, since the color-$M/L$ relations that we
use have a very shallow slope in $B-R$ color.  To examine this effect further,
for 4 galaxies that span the range of inclinations in our sample, 
we apply the \citet{tull} extinction correction
to their radial $K$-band profiles, and use their extinction-corrected 
$B-R$ colors to derive $(M/L_K)_*$.  From the
resulting surface mass-density profiles, we derive
baryonic mass rotation curves.
In Figure~\ref{fig:T98_rcs}, we plot for the 4 galaxies the difference between the 
extinction-corrected baryonic rotation curve and the uncorrected one. 
For NGC\,157, NGC\,3726, NGC\,7217, and NGC\,7606, the
rotation curves on average differ by 1.3, 1.1, 2.7, and 5.0 km s$^{-1}$, respectively.
The greatest difference is for NGC\,7606 which has the
largest inclination of the four galaxies, $63.9\degr$.
These differences in velocity are all less than the 
uncertainties of the baryonic rotation curves themselves, 
as discussed in $\S$3.3.

Galaxy disks have been observed to have a relatively constant face-on
dust opacity of $\sim 0.5$ magnitude in the $I$-band \citep{holw}.  
Correcting for this, $A_I$ would increase the baryonic velocity by $\sim 20$ km s$^{-1}$.
However, since the maximal disk limits in Figure 4 would also need to be corrected,
and the color-$M/L$ relations re-scaled accordingly, the net effect would
more or less cancel out.
However, it is very likely that more complex dust models 
where the effects of dust depend on star/dust geometry
are necessary, such as those by \citet{disn} and \citet{gord}.  
We choose to leave this
approach to a future analysis, since the detailed radiative transfer
codes are not publicly available and are difficult to apply to real
galaxies.  Moreover, different galaxies may have various amounts 
of dust and different star/dust geometries which would
cause dust to affect the baryonic rotation curve of each galaxy
uniquely.

\section{Dark Matter}
\label{sec:darkmatter}
For each galaxy, we compare its baryonic
rotation curve derived in $\S$3.2
with its observed rotation
curve taken from the literature to derive a ``dark matter
rotation curve.''

\subsection{Observed Rotation Curves}
\label{sec:observedrcs}
Observed rotation curves are plotted in Figure 3;
Table 1 lists the tracer and literature reference for each.
Some galaxies have two observed rotation curves, one from H$\alpha$ or \ion{N}{2} observations
which traces the inner parts, and another from \ion{H}{1} observations which
traces the outer parts.  If errors for the observed rotation curves were given in the original
table or plot that they were taken from, then these
are used in this paper.  If no errors are given, then we estimate
them to be the difference in rotation velocity between the approaching and 
receding sides of the galaxy; rotation curves for which we estimate the errors in this manner are
noted in Table 1.
This estimate will generally give errors larger than
the true measurement errors since it will be more affected by non-axisymmetric
features such as spiral arms and slight warps in the gas distributions.
Note that the 2--3 innermost points of the \ion{H}{1} rotation curves
and the outer few
points of the H$\alpha$ and \ion{N}{2} rotation curves have a greater uncertainty
than other points due to beam smearing and low signal-to-noise,
respectively.  For many of the rotation curves we have obtained
data from the authors, but for a handful we could not.
For those few galaxies, rotation curves are extracted
from plots in the literature with the DataThief program \citep{tumm};
these galaxies are noted in Table 1. Errors inherent to
the extraction of a rotation curve vary from plot to plot,
but tend to be $\le 5$ km s$^{-1}$.
 
Galaxies marked with the reference ``Mathewson \& Courteau'' in 
Table 1 are rotation curves that were
originally presented in \citet{math} and were later modeled by
\citet{cour}.  For these galaxies, we plot both the actual 
and model rotation curves in Figure 3.  
Since we are primarily interested
in large-scale trends, we adopt the model rotation curves
in the following analysis
in order to avoid much of the fine structure inherent to the actual
data.

\subsection{Dark Matter Rotation Curves}
\label{sec:dmrcs}
At those radii where the observed rotation speed of a galaxy is greater than that
of its baryons, the additional gravitational
component is assumed to be due to dark matter.
A dark matter rotation curve is derived as the square root of the difference of
the squares of the observed rotation velocity and the baryonic rotation curve velocity
at each radius \citep{binn}.  In doing this, it is assumed that the halos
of galaxies are axially symmetric, the disk and halo are aligned, and the observed 
gas is in circular orbits.  Dark matter rotation curves for the galaxies 
are plotted in Figure 3.

For 10 galaxies, the baryonic rotation curves over-predict the observed
rotation curves for some range in radius\footnote{We do not include 
in this count baryonic rotation curves that over-predict the inner parts of 
\ion{H}{1} rotation curves, since they are affected by beam-smearing
in this region.  Also, Figure~\ref{fig:BdeJ_upperlim} suggests that this 
number is less than 10, but we
ignore the very inner parts of the galaxies when calculating
maximal $(M/L_K)_*$s in $\S$3.1.}.
These galaxies are NGC\,157, NGC\,1559, NGC\,2139, NGC\,2841, NGC\,3198, 
NGC\,4138, NGC\,4698, NGC\,5371, NGC\,6300, and NGC\,7083. 
The baryonic rotation curve for the -0.1 dex scatter in the 
color-$M/L$ relations also over-predicts for the first six galaxies listed above.
However, for these six galaxies, the over-prediction occurs only
in the very inner parts where the observed rotation curve is affected
by structures such as rings, bars, inner windings of spiral arms,
and/or irregular morphology.  Baryonic rotation curves are derived under the
assumption of circular motion and need not trace such structures.
For NGC\,6300 and NGC\,7083, the baryonic
rotation curves for -0.1 dex do not over-predict the observed rotation curves.
The galaxy NGC\,6300 has a bar and a ring in the region where
the over-prediction occurs, but 
the image of NGC\,7083 shows no sign that its baryonic 
rotation curve should over-predict.
This is somewhat acceptable, though, since the baryonic
rotation curve for NGC\,7083 for -0.1 dex does not over-predict.

\begin{figure}[center]
\figurenum{8}
\hspace{-0.2in}
\includegraphics[height=1.3in, width=3.4in]{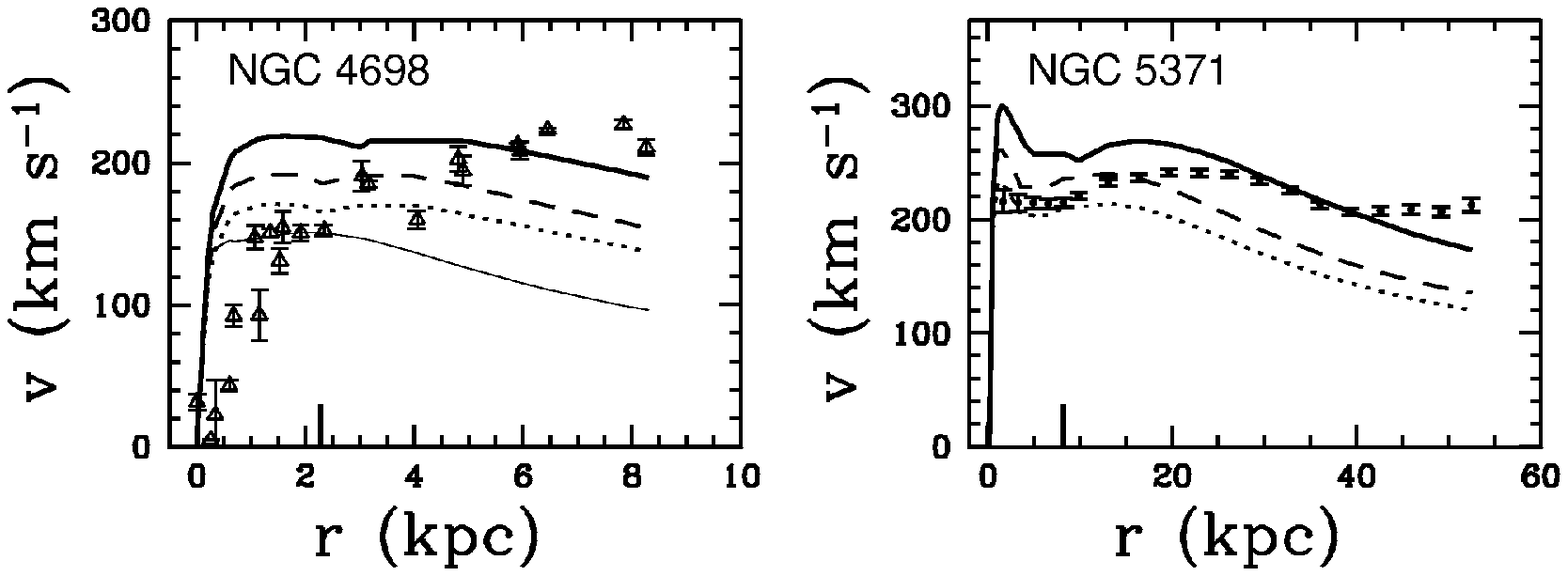}
\vspace{-0.2in}
\caption{{\small Effects on the baryonic rotation curves
of the -0.1 dex scatter in the color-$M/L$ relations 
and the -20\% uncertainty in distance for galaxies where the baryons
over-predict the observed rotation curve.  
Original baryonic rotation curves (thick solid lines), those
derived by taking into account the
distance uncertainty (dashed lines), and those derived by taking into
account both the distance uncertainty and the color-$M/L$
scatter (dotted lines) are triangles for H$\alpha$ and small
points for \ion{H}{1}.
For NGC\,4698, we also plot the baryonic rotation curve derived 
from the adoption of a different distance to the galaxy (thin solid line).  
Observed rotation curves are plotted as
points.  The radius $2.2h_{IR}$ is marked as a thick
solid line on the x-axis.}
\label{fig:diff_dist}}
\end{figure}

Two galaxies, NGC\,4698 and NGC\,5371, have baryonic rotation curves that
over-predict their observed rotation curves, even for the -0.1 dex scatter
in the color-$M/L$ relations.
For both these galaxies, the over-prediction cannot only be explained by bars
and/or rings in the galaxy images.  To examine things further, 
we create baryonic rotation curves using the position angle and 
inclination that were used to derive their
observed rotation curves; the resulting curves also 
over-predict.  Next, we create baryonic rotation curves 
for the -20\% uncertainty in the galaxies' distances and both the
distance and color-$M/L$ uncertainties; they are plotted
in Figure~\ref{fig:diff_dist}.
For NGC\,4698, the baryonic rotation
curve created by taking into account both these effects still over-predicts.
However, for this galaxy, its distance was calculated with a
Virgocentric infall calculation and was found to be triple-valued (Paper\,I).
We derive a baryonic rotation curve for it with the distance solution
which is closer than the one chosen in Paper\,I (9.7 Mpc, as opposed to
the chosen solution of 19.1 Mpc).  This baryonic rotation curve
is plotted in Figure~\ref{fig:diff_dist} and
over-predicts only the inner portion ($\sim 1.5$ kpc) of the observed 
rotation curve.  For NGC\,5371, the baryonic rotation curve created by taking into
account the distance uncertainty over-predicts, while that
created by taking into account both distance and color-$M/L$ uncertainties
does not.
The sort of over-prediction that is observed for NGC\,4698 and NGC\,5371
can also be due to such factors as a mis-measurement of 
the observed rotation curves and/or the effects of dust.
A stellar population affected by dust may appear redder (and hence heavier and
have a faster rotation) than it is intrinsically since the reddening 
effect of dust is slightly greater than its extinction effect.
These galaxies could also signal a need to lower the normalization of
the color-$M/L$ relations.  If we did this, then many of the
other galaxies in the sample would be submaximal, but the 
qualitative results of this paper would not change.

\subsection{Uncertainties in Dark Matter Rotation Curves}
Uncertainties in the dark matter rotation curves arise from a number of
effects. The most significant are those 
inherent to the determination of the baryonic mass component,
as discussed in $\S$3.3. Other uncertainties 
include: non-circular motions that perturb the underlying potential
(i.e., spiral arms, bars, substructure), statistical errors from
the measurement of velocities in radial bins, systematic errors in
measuring the velocity (i.e., beam smearing and slit position angle
error), and uncertainties in the measurement of the dynamical centers
of the galaxies.  There also may be differences between the centers of galaxies
determined from photometry and those determined from the observed rotation
curves.  However, center measurements are not expected to differ much since
photometric centers are always chosen to be the brightest pixel
in the nucleus which coincides with the dynamical center of most galaxies.

\section{Maximal Disks}
The radius $R=2.2 h$ is where the rotation curve of a self-gravitating
exponential disk reaches its peak \citep{free}.  A commonly used definition of a maximal
disk is given by \citet{sack} where the galaxy $disk$ provides $85\% \pm 10\%$
of the total rotational support of the galaxy at $2.2h_R$.  In the following analysis, 
we define a galaxy to have a maximal disk if it has a baryonic mass (disk and bulge)
contribution to the observed rotation curve of $>90\%$ at $2.2h$, or similarly,
a dark matter contribution to the observed rotation curve of $<10\%$ at $2.2h$.  We choose to adopt
this definition of maximal disk over that of \citet{sack} since we perform
detailed bulge-disk decompositions, and are therefore able to model the combined
bulge plus disk baryonic rotation curves.  Furthermore, we use $h$ 
measured at $K$, and when $K$-band imaging is unavailable,
we use $H$.  Since near-infrared bands trace most of the mass of the stellar populations,
a near-infrared scale-length is more analogous to the stellar mass scale-length of a 
disk.  Disk scale-lengths measured at $K$
are typically $\sim 1.2$ times shorter than those measured at $B$ \citep{dej2}.

In Figure 3, there are 4 galaxies that have submaximal disks,
even if the -0.1 dex scatter in the color-$M/L$ relations
is taken into account.  Since the normalization of the color-$M/L$ relations is an upper limit,
the number of galaxies in our sample that do not have maximal disks can 
in principle be much greater than 4.
However, for those galaxies that 
we observe to have maximal disks, the overall shapes\footnote{This is distinguished
from the ``bumps and wiggles'' of the observed rotation curves that are likely 
due to small-scale features such as bars and
spiral arm perturbations (\citealt{palu,kslyz,slyzk}).}  
of the inner parts of their observed rotation curves (within $\sim R_{25}$) are generally 
matched by the shapes of their baryonic rotation curves.
This is evidence that many of the disks have a significant baryonic component
in their inner parts.  

\begin{figure}[center]
\figurenum{9}
\hspace{-0.3in}
\includegraphics[height=4in, width=3.6in]{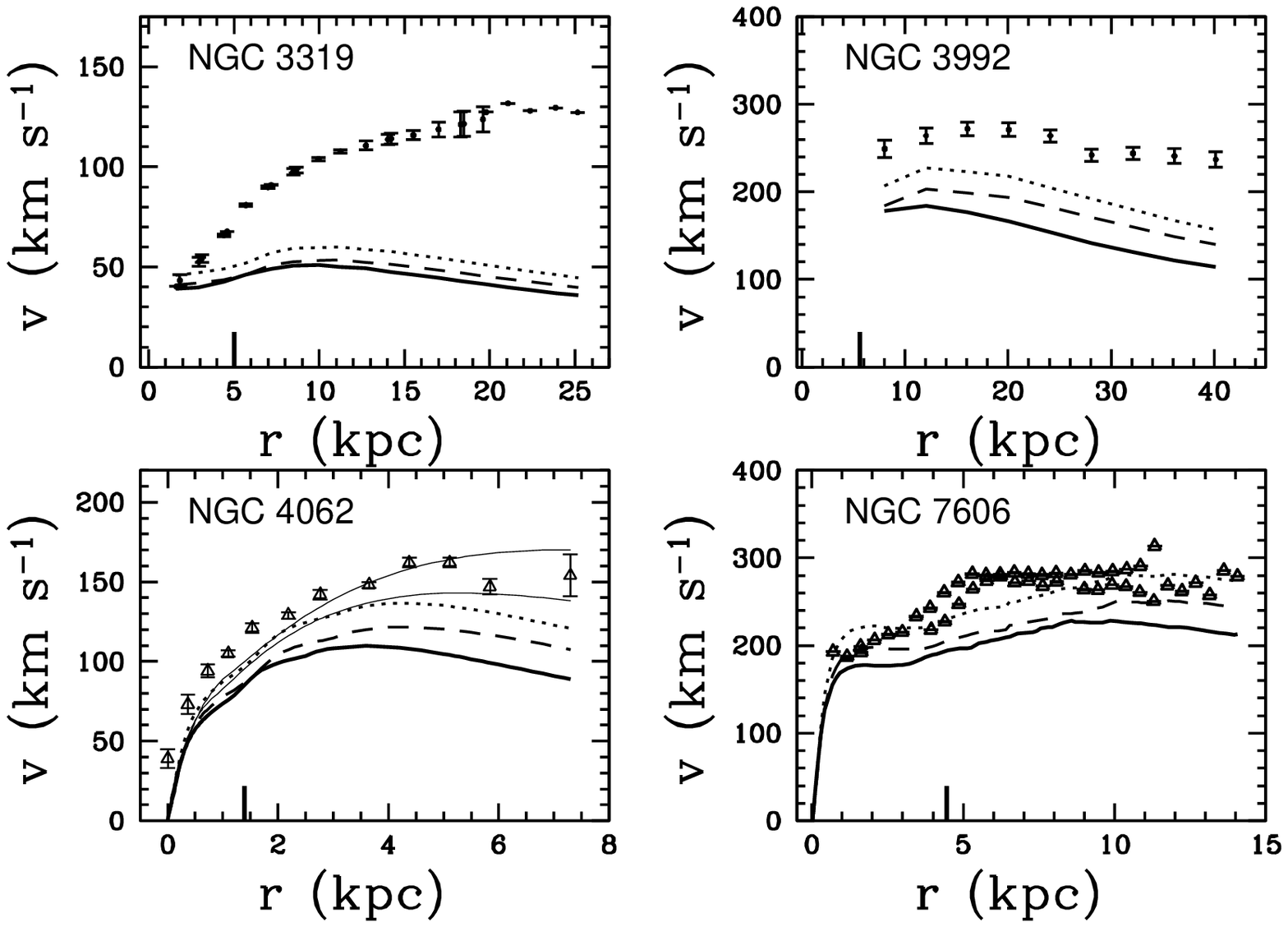}
\vspace{-1.5in}
\caption{{\small Same as Figure~\ref{fig:diff_dist}, but for galaxies where the
baryons under-predict the observed rotation curves, and for the +0.1 dex
scatter in the color-$M/L$ relations and the +20\% distance uncertainty. 
For NGC\,4062, baryonic rotation curves derived for
2 different distance determinations are plotted
as thin solid lines.}
\label{fig:diff_dist2}}
\end{figure}

These 4 galaxies are NGC\,3319\footnote{For NGC\,3319, we do not take into account its innermost
2 observed rotation curve points since they are measured from \ion{H}{1} and
are likely affected by beam-smearing.  Also, the
distance measurement for NGC\,3319 is from Cepheid variable stars
and has an uncertainty of only $\sim 10\%$.}, NGC\,3992,
NGC\,4062, and NGC\,7606; they have baryonic mass contributions to their
total masses at $2.2 h_{IR}$ of 22\%, 50\%, 57\%, and 68\%, respectively.
In Figure~\ref{fig:diff_dist2} we plot for these 4 galaxies
the baryonic rotation curves with and without uncertainties in distances
and the color-$M/L$ relations taken into account; these
baryonic rotation curves still under-predict the observed rotation curves
for NGC\,3319, NGC\,3992, and NGC\,4062.
However, for NGC\,7606, the baryonic rotation curve that takes into account
both these effects follows the observed rotation curve fairly well. 
For NGC\,4062, its distance was calculated using a Virgocentric infall
calculation and was found to be triple-valued (Paper\,I).  We create
baryonic rotation curves for NGC\,4062 with the other 2 distance solutions (17.6 and 24.4 Mpc),
both of which are greater than the one chosen in  Paper\,I; they are
plotted in Figure~\ref{fig:diff_dist2}.
These baryonic rotation curves have a strange behavior since they under-predict
the observed rotation curve in the inner parts, but match it in the outer regions.
This may be the effect of such factors as an underestimate of the stellar mass in the inner parts 
of the galaxy due to blue spiral arms in the $B$-band image (see image in Figure 1 of Paper\,I),
a poor bulge/disk decomposition, and/or an underestimate of the 
uncertainties of the observed rotation curve.  In summary, while NGC\,4062 and 
NGC\,7606 may not have submaximal disks, NGC\,3319 and NGC\,3992 likely do.

There are a few galaxies that have marginally submaximal disks.  For these
galaxies, the baryonic rotation curves for the +0.1 dex scatter in the
color-$M/L$ relations result in maximal disks. These galaxies are
NGC\,1241, NGC\,2139, and NGC\,2280, and they have baryonic mass contributions
to their total rotational support at $2.2 h_{IR}$ of 75\%, 73\%, and 72\%, respectively.  

One striking example of a galaxy that is likely close to maximum disk
is NGC\,157.  This galaxy has H$\alpha$ and \ion{H}{1} rotation curves 
that have a sudden steep decline at
$\sim 8$kpc ($\sim 3 h_K$) and flatten afterward, which results in a 
hump-like structure.
While this peculiar behavior cannot be absolutely confirmed by \citet{ryd2}, there 
are strong lines of evidence presented in their paper that point to
this hump-like structure as physical.  This same structure is also
found in the baryonic rotation curve for NGC\,157.  Moreover, a dark matter halo
in the shape of a NFW model is not consistent with this structure unless baryons make
a significant contribution to the inner $\sim 5$ scale-lengths of NGC\,157.
In summary, although a lower normalization of the IMF cannot be ruled out,
it seems unlikely that at least NGC\,157 is strongly submaximal.

\section{Dark and Baryonic Matter Scaling Relations}
In this section, we examine scaling relations for dark and baryonic matter.  We use 
quantities from Tables 1 and 4 of Paper\,I, and
derive others: $V_{tot,max}$ the maximum
observed rotation curve velocity, $V_{b,max}$ the maximum baryonic
rotation curve velocity, $R(V_{b,max})$ the radius at which $V=V_{b,max}$,
and $M_b$ the baryonic mass.  
These quantities are listed in Table 2, and if they are derived from the 
baryonic mass distributions, we also tabulate the differences in their
values for renormalizations of the color-$M/L$ relations 
by $\pm0.1$ and -0.3 dex.
We choose to create the quantity $V_{b,max}$ because it 
can be derived from imaging alone and obviates the need for much more 
expensive and time-consuming line width observations needed to obtain
$V_{tot,max}$.  For bright spirals, $V_{b,max}$ should not differ from $V_{tot,max}$ by very
much, and due to the flat and usually noisier nature of the observed rotation
curves, the measurement of $V_{b,max}$ is more straightforward than
that of $V_{tot,max}$. This quantity will be discussed further in $\S$6.2.

We derive two quantities from the dark matter rotation curves calculated
in $\S$4.2: $R_{10}$,
the radius where dark matter contributes $10\%$ to the velocity of the observed
rotation curve, and $R_X$, the radius where the dark matter
contribution equals that of the baryons (the ``cross-over radius'').
The quantity $R_{10}$ is similar to $R_{\rm IBD}$ of \citet{salu} 
and $R_{\rm t}$ of Persic et al.\ (1996).  The radius $R_X$ is
analogous to $R_{2:1}$ of \citet{mc98}.  The radii
$R_{10}$ and $R_X$ 
are listed in Table 2 along with the difference in values for
renormalizations of the color-$M/L$ relations by $\pm0.1$ and -0.3 dex.  
Note that for some galaxies we cannot measure $R_{10}$ or $R_X$; 
this is in general because their rotation curves do not extend far enough
in radius.

\subsection{Baryonic Scaling Relations}
\begin{figure}[center]
\figurenum{10}
\hspace{-0.25in}
\includegraphics[height=3.6in, width=3.35in]{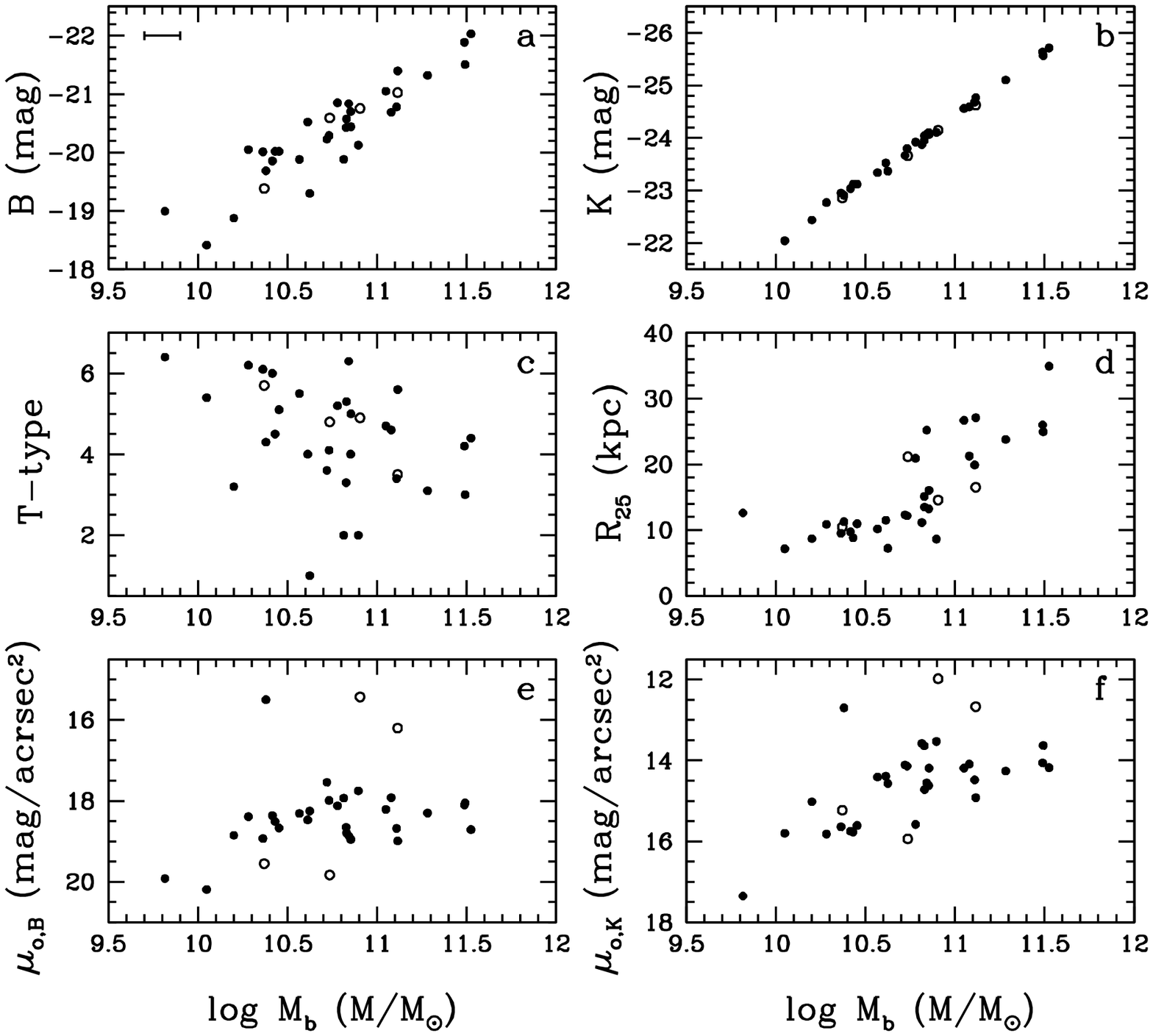}
\vspace{-0.5in}
\caption{{\small Basic physical parameters
  of galaxies versus baryonic mass.  Open circles represent galaxies from
  the SDSS that have partial imaging. A typical error bar for 
  $M_b$ is plotted in panel a; error bars for the other quantities 
  are the size of the data points at the resolution of the figure.
  To make the plot in panel c such that
  points do not overlap, we add fractions with values $<1$ to the
  integer Hubble T-types.} \label{fig:reltns_htype}}
\end{figure}

In Figure~\ref{fig:reltns_htype}, basic physical parameters of galaxies
are plotted versus $M_b$: total absolute $B$ and $K$-band magnitudes,
Hubble T-type, $R_{25}$, and $B$ and $K$-band central surface brightnesses ($\mu_{o,B}$, $\mu_{o,K}$).
They have the following correlation coefficients: 0.91 ($B$), 1.00 ($K$),
0.40 (Hubble T-type), 0.78 ($R_{25}$), 0.29 ($\mu_{o,B}$), and 0.58 ($\mu_{o,K}$).
Integrated magnitudes, sizes, $\mu_{o,K}$, and maximum rotation velocities
correlate very well with $M_b$, such that
galaxies with greater $M_b$ are brighter, larger, and have brighter 
$\mu_{o,K}$.  Morphology and $\mu_{o,B}$ also correlate with $M_b$, but to
a lesser extent.  

\begin{figure}[center]
\figurenum{11}
\hspace{-0.3in}
\includegraphics[height=3.4in, width=3.4in]{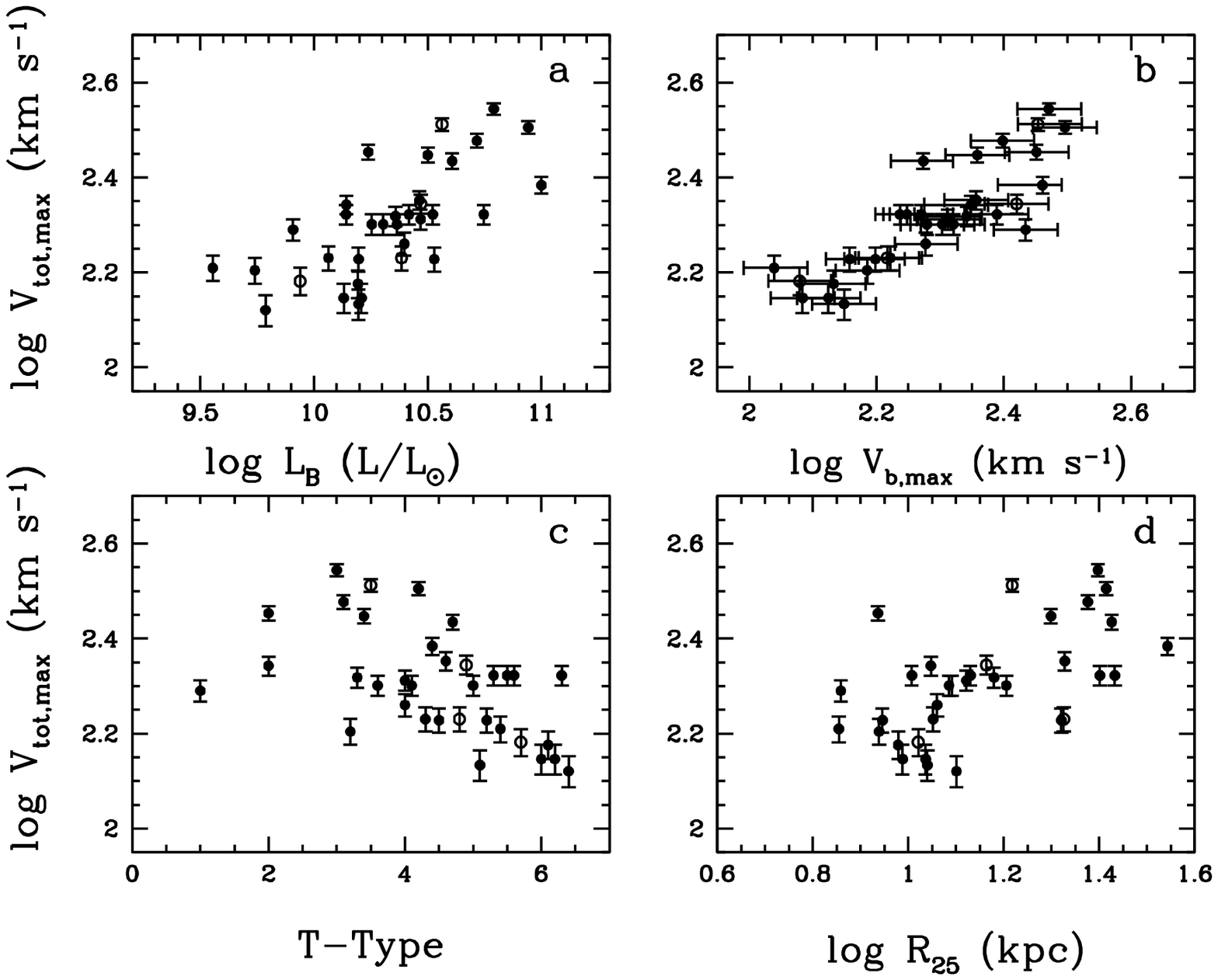}
\vspace{-1.in}
\caption{{\small Relations for $V_{tot,max}$
  and $V_{b,max}$, which trace the
  total and baryonic mass component of the galactic
  potential, respectively.  Open circles represent galaxies from
  the SDSS that have partial imaging.  Error bars for T-type,
$L_B$, and $R_{25}$ are the size of the data points at the
resolution of the figure.}  
\label{fig:reltns_v}}
\end{figure}

In Figure~\ref{fig:reltns_v}, relations are plotted for log$_{10} V_{b,max}$
and log$_{10} V_{tot,max}$, which trace the baryonic and total
mass components of the galactic potential, respectively. 
In Figure~\ref{fig:reltns_v}a, along the lines of \citet{robe}, we 
plot log$_{10} V_{tot,max}$ versus log$_{10} L_B$ and find
good agreement with their results.  In Figure~\ref{fig:reltns_v}b, we plot 
log$_{10} V_{b,max}$ versus log$_{10} V_{tot,max}$; they are strongly correlated with a correlation
coefficient of 0.82.  These velocities should be equivalent for maximal disks,
which many of the galaxies in our sample have.  In addition, a tight correlation
indicates that there is not a wide spread in the degree of maximality.
The quantity log$_{10} V_{tot,max}$ also correlates with Hubble T-type and $R_{25}$
such that galaxies that rotate faster have earlier T-types and
are larger.  The quantity $R(V_{b,max})$ correlates with log$_{10} M_b$ with
a correlation coefficient of 0.66, but correlates relatively weakly
with log$_{10} V_{b,max}$, log$_{10} V_{tot,max}$, and $\mu_{o,K}$ with
correlation coefficients of 0.08, 0.35, and 0.16, respectively.

\begin{figure}[center]
\figurenum{12}
\includegraphics[height=3.0in, width=3.0in]{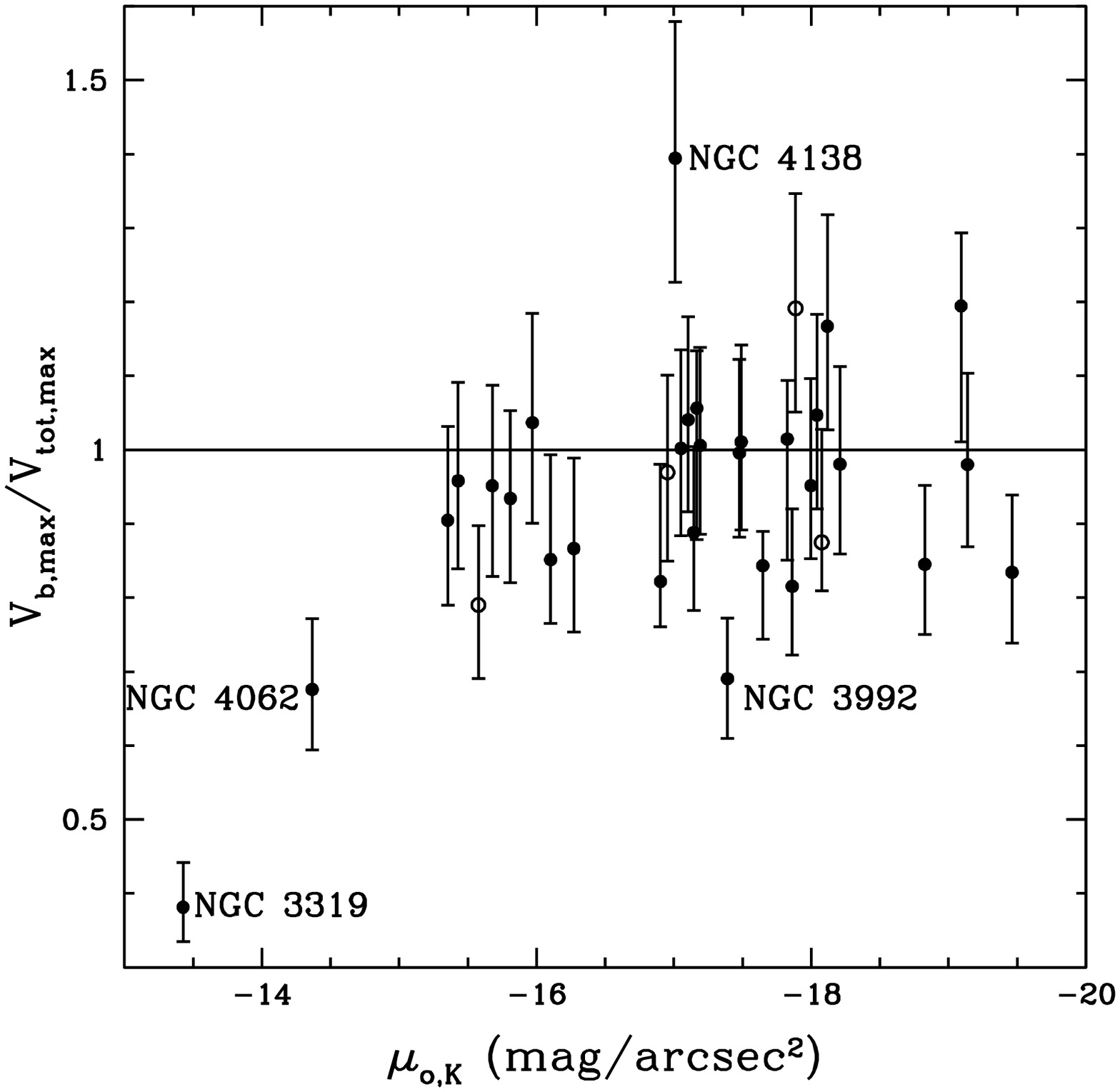}
\caption{{\small Ratio of maximum baryonic velocity to maximum observed velocity versus
$\mu_{o,K}$.  SDSS galaxies with partial imaging are
plotted as open circles.  Outliers are labeled: NGC\,4062 and NGC\,3992
have submaximal disks, NGC\,4138 has evidence of kinematic disturbance,
and NGC\,3319 has both a submaximal disk and a kinematic disturbance.}
\label{fig:ratio_vels}}
\end{figure}

As discussed above, the maximum rotation speeds predicted from the baryon
distributions are tightly correlated with the observed maximum rotation speeds.
To examine this relation further,
we plot the ratio of $V_{b,max}$ to $V_{tot,max}$ versus $\mu_{o,K}$ in 
Figure~\ref{fig:ratio_vels}, and a find root mean square (rms) deviation 
from unity of only 0.18.  Using this result, one can create a Tully-Fisher relation from two
passband surface photometry and a redshift alone (e.g., by using the SDSS and the color-$M/L$
relations for $g-r$ and $(M/L_r)_*$).  The 4 outliers in Figure~\ref{fig:ratio_vels} are galaxies 
that have submaximal disks and/or are kinematically disturbed.  The relation between
$V_{b,max}$ and $V_{tot,max}$ is not expected to be as tight for less luminous spirals,
but could possibly be calibrated for such a population.
\begin{figure}[center]
\figurenum{13}
\hspace{-0.3in}
\includegraphics[height=3.8in, width=3.4in]{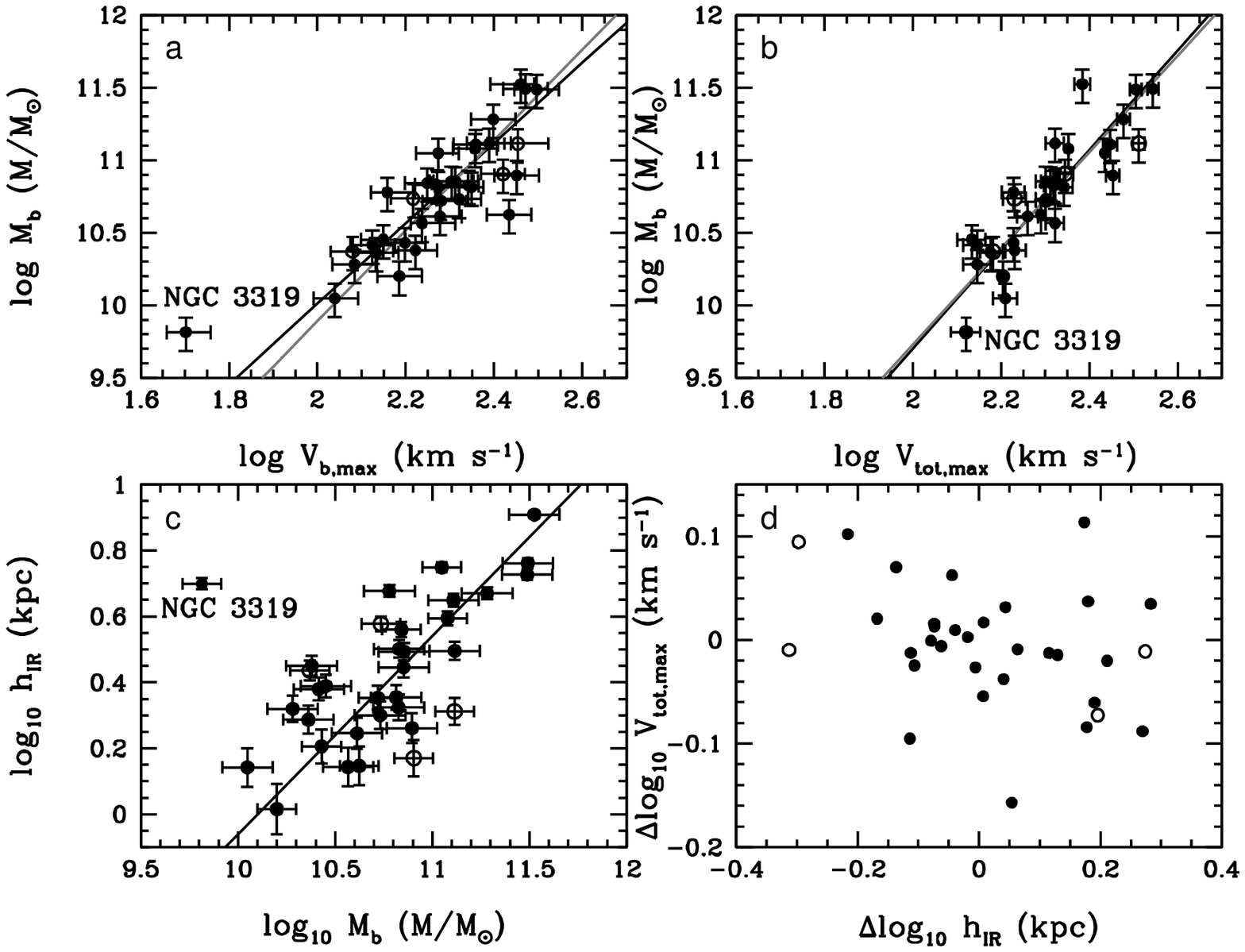}
\vspace{-1.0in}
\caption{{\small Baryonic Tully-Fisher relations for
 $V_{tot,max}$ and $V_{b,max}$, and the size-baryonic mass relation are plotted 
 in panels a--c.  Weighted bisector fits with and without
 NGC\,3319 are plotted as black and gray lines, respectively.  
 In panel d, the $V_{tot,max}$ residual of the Tully-Fisher relation
 is plotted versus the size residual from the size-baryonic mass
 relation. Galaxies from the SDSS that have partial imaging are
 plotted as open circles.}
\label{fig:reltns_tf}}
\end{figure}

In Figure~\ref{fig:reltns_tf}a,b we plot the baryonic Tully-Fisher 
relations for $V_{tot,max}$ and $V_{b,max}$.  Weighted bisector fits 
are given in the form: $M_b = M_{200} V^{\alpha}$ where $M_{200}$ is the 
stellar mass in units of $10^{10} \rm M_{\odot}$ for a galaxy on the Tully-Fisher 
relation with $V=200$ km s$^{-1}$.  They are: $M_{200}=5.40 \pm 0.03$ and
$\alpha=3.4 \pm 0.3$ with a rms residual of 0.20 for $V_{tot,max}$,
and $M_{200}=6.98 \pm 0.03$ and $\alpha=2.8 \pm 0.3$ with a rms
residual of 0.22 for $V_{b,max}$.  If we do not include the outlier in 
these two relations, NGC\,3319, we obtain the following fits:
$M_{200}=5.36 \pm 0.03$ and $\alpha=3.3 \pm 0.3$ with a rms residual 
of 0.19 for $V_{tot,max}$, and $M_{200}=6.75 \pm 0.03$ and $\alpha=3.1 \pm 0.3$ 
with a rms residual of 0.21 for $V_{b,max}$.  
The galaxy NGC\,3319 is both
kinematically disturbed on one side \citep{moor}, and has the smallest stellar
mass in our sample.
For most of the galaxies in our sample, we do not have
\ion{H}{1} or molecular hydrogen measurements.  If gas
masses were included, the Tully-Fisher slope of $M_b$ on $V_{tot,max}$ would 
flatten, as lower mass galaxies (e.g., NGC\,3319) have a larger gas content
than more massive galaxies \citep[e.g.,][]{verh}.  

We refrain from comparing the slopes found here and in the literature
\citep[e.g.,][]{bdej,mcg5,piza}
for the following reasons.  Such fits are sensitive, among other things, to 
the velocity measurements used (i.e., $V_{tot,max}, V_{b,max}, V_{flat}, V_{2.2}$, $W_{20}$),
the range in stellar mass of the samples, and the galaxy samples used (range in
Hubble type, field versus cluster environments).  In particular, \citet{guro}
find a break in the stellar mass and baryonic Tully-Fisher
relations for less massive galaxies ($10^{7.6} \lesssim M_* \lesssim 10^{9.6} M_{\odot}$).
This break is such that the Tully-Fisher relation for less massive galaxies
is steeper than that for galaxies with stellar masses from
$\sim 10^{9.6}$--$10^{11.2} M_{\odot}$.  \citet{guro} find slopes for
the lower and upper mass ranges of their $H$-band sample, using $W_{20}$ as
a velocity measurement, of $4.4\pm0.3$ and $3.3\pm0.3$, respectively.
This break/curvature makes a comparison between slopes of different samples
risky at best, especially when different selection criteria are used.
A 2D Kolomogrov-Smirnov test would be ideal, but beyond the scope of this
work considering the modest stellar mass range of our sample.

The size-baryonic mass relation, where size is parameterized by $h_{IR}$,
is plotted in Figure~\ref{fig:reltns_tf}c.  We perform a weighted bisector
fit and find $h_{IR}= {0.67 \pm 2.31}\, (M_b/10^{10} M_{\odot})^{0.60 \pm 0.08}$ kpc
with a rms of 0.16.  We compare the $V_{tot,max}$ residuals of the
Tully-Fisher relation with the size residuals of the size-mass relation
in Figure~\ref{fig:reltns_tf}d, and find a scatter plot. 
If galaxy disks are maximal and the dark matter halos of galaxies of
different surface brightness are identical, then it is predicted by \citet{mc98} and 
\citet{crix} that there should be
a correlation between these residuals.  \citet{crix} argue that the lack
of residual correlation, essentially the surface brightness independence
of the Tully-Fisher relation, implies that all disks are submaximal. 
This argument, however, is not entirely 
straightforward, since in the \citet{piza} model for disk collapse within a halo, variations in dark 
matter halo parameters can create enough scatter
in the Tully-Fisher relation to hide the predicted correlation.
Furthermore, \citet{mc98} and \citet{sell}
argue that submaximal disks do not solve the problem, and one is left with a fine-tuning
problem: either disk $M/L$s correlate with surface brightness,
halo contributions vary, or Newtonian dynamics falters.
Figure 12 illustrates this fine-tuning problem by showing that $V_{b,max}/V_{tot,max}$
is independent of surface brightness, so that the total velocity ``knows about'' the
baryonic contribution.  Changing the normalization of the color-$M/L$
relations re-scales the y-axis of the plot, but the ratio remains 
independent of surface brightness.

\subsection{Dark Matter Scaling Relations}
\begin{figure*}
\figurenum{14}
\includegraphics[scale=0.75]{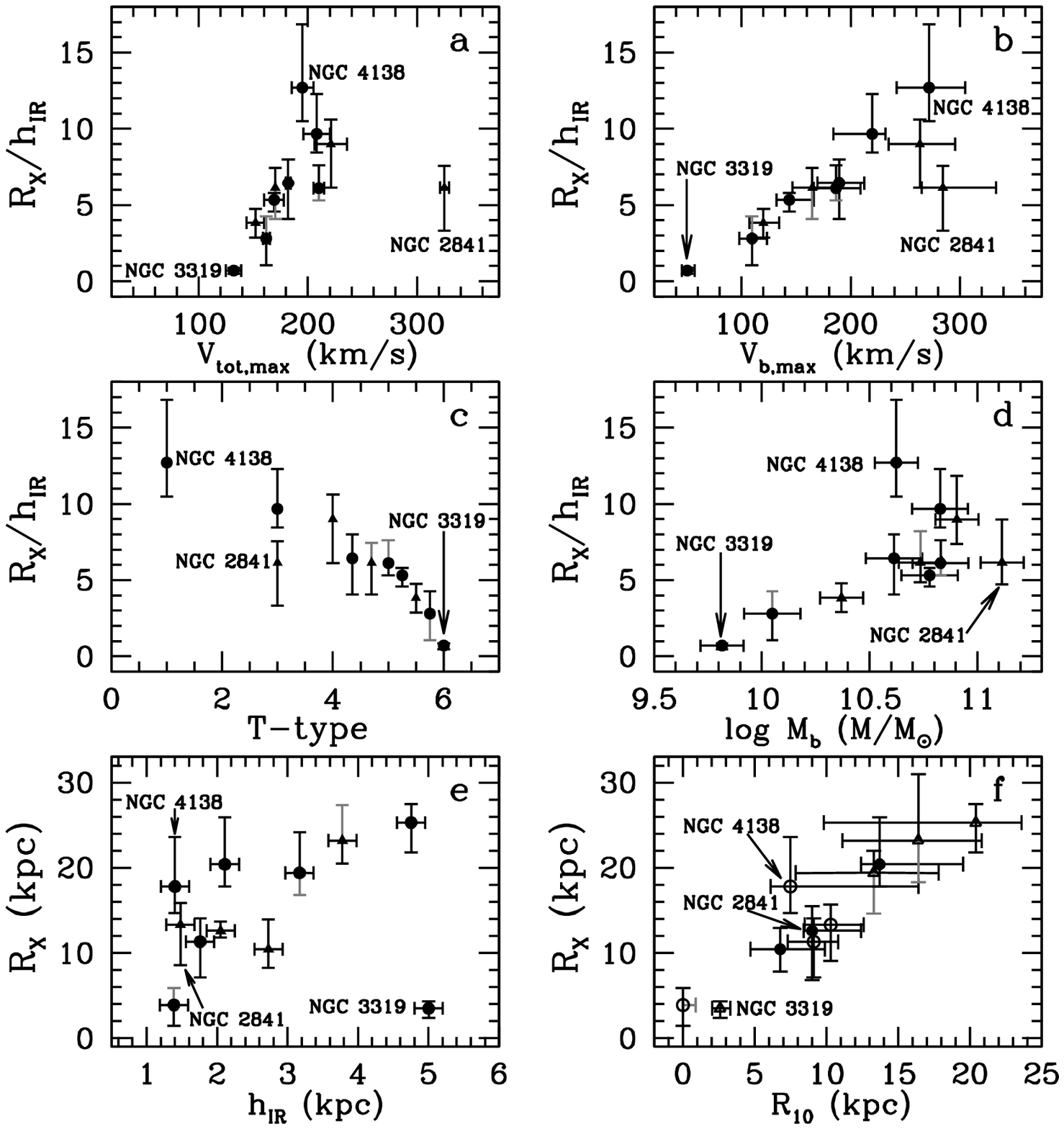}
\vspace{-0.2in}
\caption{{\small Relations for $R_X/h_{IR}$ and $R_X$. In panels a--e, triangles
represent galaxies from SDSS that only have partial imaging.
Error bars represent the values of the quantities derived 
by renormalizing the color-$M/L$ relations by $\pm 0.1$ dex;
upper and lower limits to these values are plotted in gray.
To make the points in panel c such that they do not overlap, 
we add fractions with values $<1$ to the integer Hubble types.
In panel f, galaxies are plotted with different symbols according to their
$h_{IR}$: those with $h_{IR} < 1\ \rm kpc$
as open circles, those with $1\ \rm kpc \le h_{IR} < 3\ \rm kpc$ as filled circles, and
those with $3\ \rm kpc \le h_{IR}$ as open triangles.
The galaxies NGC\,2841, NGC\,3319, and NGC\,4138 are labeled
in all panels; they have evidence of kinematical
disturbances despite their normal appearances.
The three outliers in Figures~\ref{fig:reltns_dm} and \ref{fig:reltns_dm2} are NGC\,2841, NGC\,3319,
and NGC\,4138.  Despite their normal optical morphologies, these galaxies
have signatures of kinematic disturbances.
The galaxy NGC\,2841 has a warp in its outer \ion{H}{1} disk \citep{bosm}
and an indication of a counter-rotating stellar component
for $5\arcsec \le r \le 12\arcsec$ \citep{silc}.
Similarly, NGC\,4138 has both a
counter-rotating disk and a significant warp in its outer \ion{H}{1}
disk \citep{jore}.  This warp may be the cause of the decline of its 
\ion{H}{1} rotation curve, and hence what makes this galaxy an outlier.
The third outlier, NGC\,3319, is discussed in $\S$6.1.}
\label{fig:reltns_dm}}
\end{figure*}

\begin{figure}[center]
\figurenum{15}
\hspace{-0.3in}
\includegraphics[height=3.6in, width=3.4in]{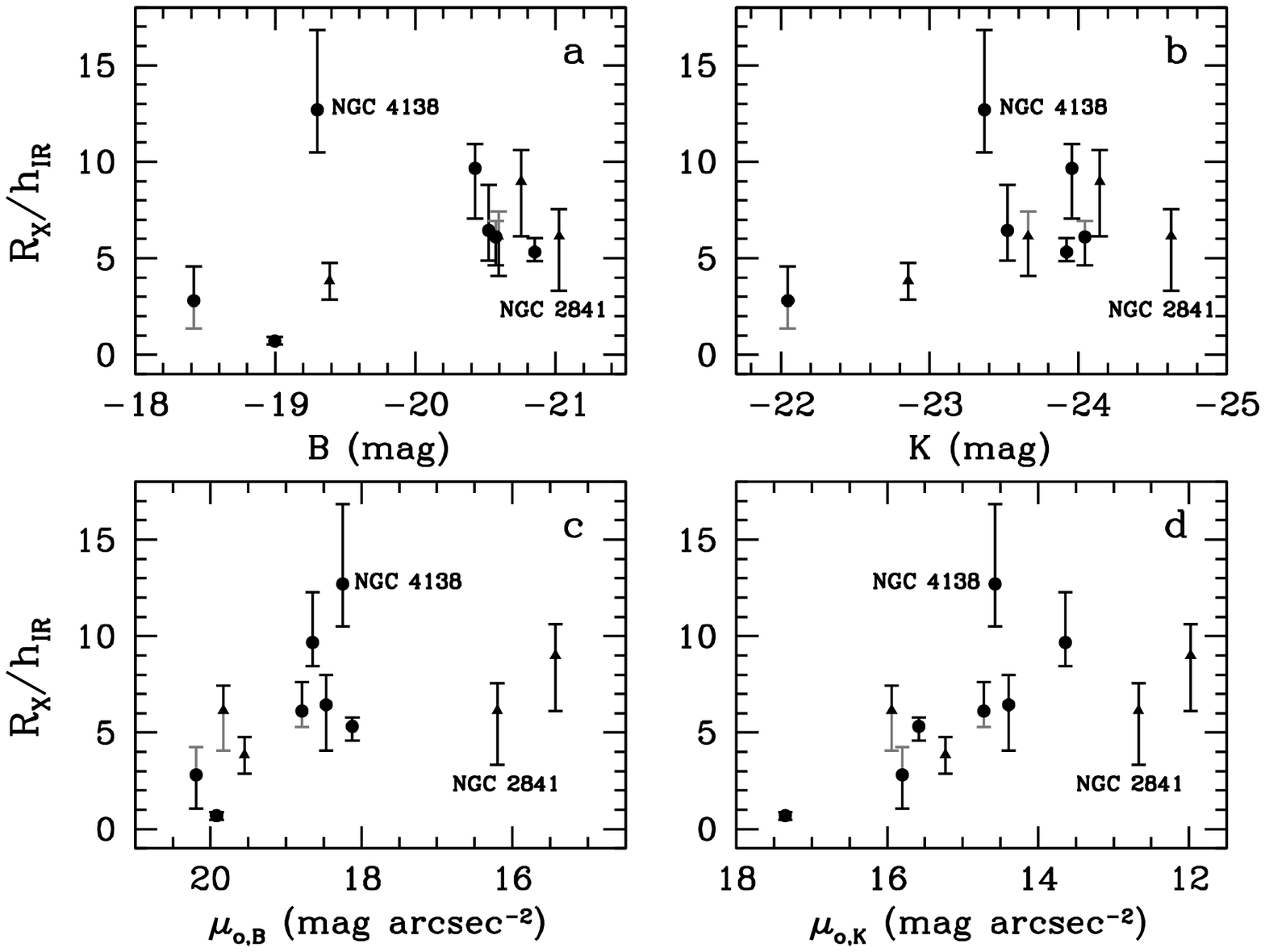}
\vspace{-1.in}
\caption{{\small Relations for $R_X/h_{IR}$ and integrated magnitudes and
central surface brightnesses.  Symbols and error bars are the same as in Figure~\ref{fig:reltns_dm}.
The galaxy NGC\,3319 has a $\mu_{o,K}$ measurement, but not a $K$ measurement.
Error bars along the x-axis are approximately the point size at the 
resolution of the plot.}
\label{fig:reltns_dm2}}
\end{figure}

In Figures~\ref{fig:reltns_dm}a--d and \ref{fig:reltns_dm2}, the baryon/dark matter
equality radius in units of the disk scale length, $R_X/h_{IR}$, 
is plotted versus a number of galactic parameters: $V_{tot,max}$, $V_{b,max}$, 
Hubble T-type, $M_b$, $B$, $K$, $\mu_{o,B}$, and $\mu_{o,K}$.
In Figure~\ref{fig:reltns_dm}e, we plot $R_X$ versus $h_{IR}$.
There are fewer points in these figures than in previous ones
since not all galaxies have a
dark matter rotation curve that allows us to determine $R_X$.
The correlation coefficients for these relations and for $R(V_{b,max})$ 
with $R_X$ are listed in Table 4.
The radius $R_X$ is found to correlate most strongly with $V_{b,max}$ and
very strongly with $M_b$, Hubble T-type, and $V_{tot,max}$.
There are no changes in the relative strengths of the relations 
if the color-$M/L$ relations are renormalized by $\pm0.1$ or $-0.3$ dex.

The trends with $R_X$ that we find are qualitatively in agreement with
those in the literature \citep[e.g.,][]{deb7,mc98,piza}, with some exceptions that are discussed here.
\citet{mc98} and \citet{deb7} find that $\mu_{o,B}$ plays a major role
in relations with galaxy properties, including with $R_{2:1}$, which
is analogous to our $R_X$ parameter.  We find that $R_X/h_{IR}$ is 
correlated with $\mu_{o,B}$, but that it is even more correlated with 
$\mu_{o,K}$.  We do not find a correlation between the mass discrepancy
and $\mu_{o,K}$ (in Figure 12 we plot the inverse mass discrepancy), 
but \citet{mc98} do find a relation between total $M/L_B$ at $4h$ and $\mu_{o,B}$.
Also, we find that $h_{IR}$ correlates strongly with $R_X$, once
outliers are removed, which is not in qualitative agreement with other studies:
\citet{mc98} and \citet{zava} find that $h$
does not correlate with total $M/L_B$ evaluated at $4h$ or the mass discrepancy
evaluated at maximum rotation velocity, respectively.
However, the results of \citet{piza} for relations with $h$ are consistent with ours;
they find that more compact galaxies have a larger mass discrepancy 
(measured at $2.2h_i$) than larger galaxies.  \citet{zava} find a number of
relations involving a quantity similar to the mass discrepancy
that are in contradiction to those presented in this paper
and in the literature.


Here we discuss possible causes of these discrepancies.
Our differences with \citet{mc98} may be traced to the 
quantities used in the analyses: they use $M/L_B$ and $h_B$, while we use $R_X$ and $h_{IR}$.
Luminosity in the $B$-band should have more scatter in relations
with galaxy properties than baryonic mass derived from a combination
of optical and near-infrared data.  This is because
the $B$-band is more affected by star formation and extinction, 
while the $K$-band is a better tracer of stellar mass.  
In addition, a near-infrared scale-length
is more analogous to the baryonic mass scale-length of a disk, and
hence should have less scatter in its relations with galaxy
properties than an optical scale-length.  Adding some credence to the hypothesis
that different quantities are the root of the discrepancy,
\citet{mc98} find that relations between $R_{2:1}$
and both $\mu_{o,B}$ and $M_B$ are such
that brighter galaxies have larger values of $R_{2:1}$, consistent with our results.
That \citet{mc98} find a correlation between $M/L_B$
evaluated at $4h$ and $\mu_{o,B}$ is likely also due to the
larger range in surface brightness in their sample.

In Figure~\ref{fig:reltns_dm}f, $R_{10}$ is plotted versus $R_X$,
and galaxies are plotted as different symbols according to their value of $h_{IR}$.
Although there is some scatter, as $h_{IR}$ increases, these two radii
move further out in the disks in tandem.  
That there is such a tight relation between $R_{10}$ and $R_X$ 
tells us that
dark matter contributions to the observed rotation curves must
increase in a characteristic way between these two radii for all
galaxies.  This is likely due to the combined effects of quasi-exponential
disks and observed rotation curves that are nearly flat. 
If this is correct, then $R_{10}$ should have more scatter
in its relations with galaxy properties than $R_X$ because the
observed rotation curves should not yet be flat in the region
where $R_{10}$ is measured, which is found to be the case.

\section{Radial Behavior of Dark Matter}
\begin{figure}[center]
\vspace{-0.5in}
\figurenum{16}
\hspace{-0.5in}
\includegraphics[height=4.3in, width=4.3in]{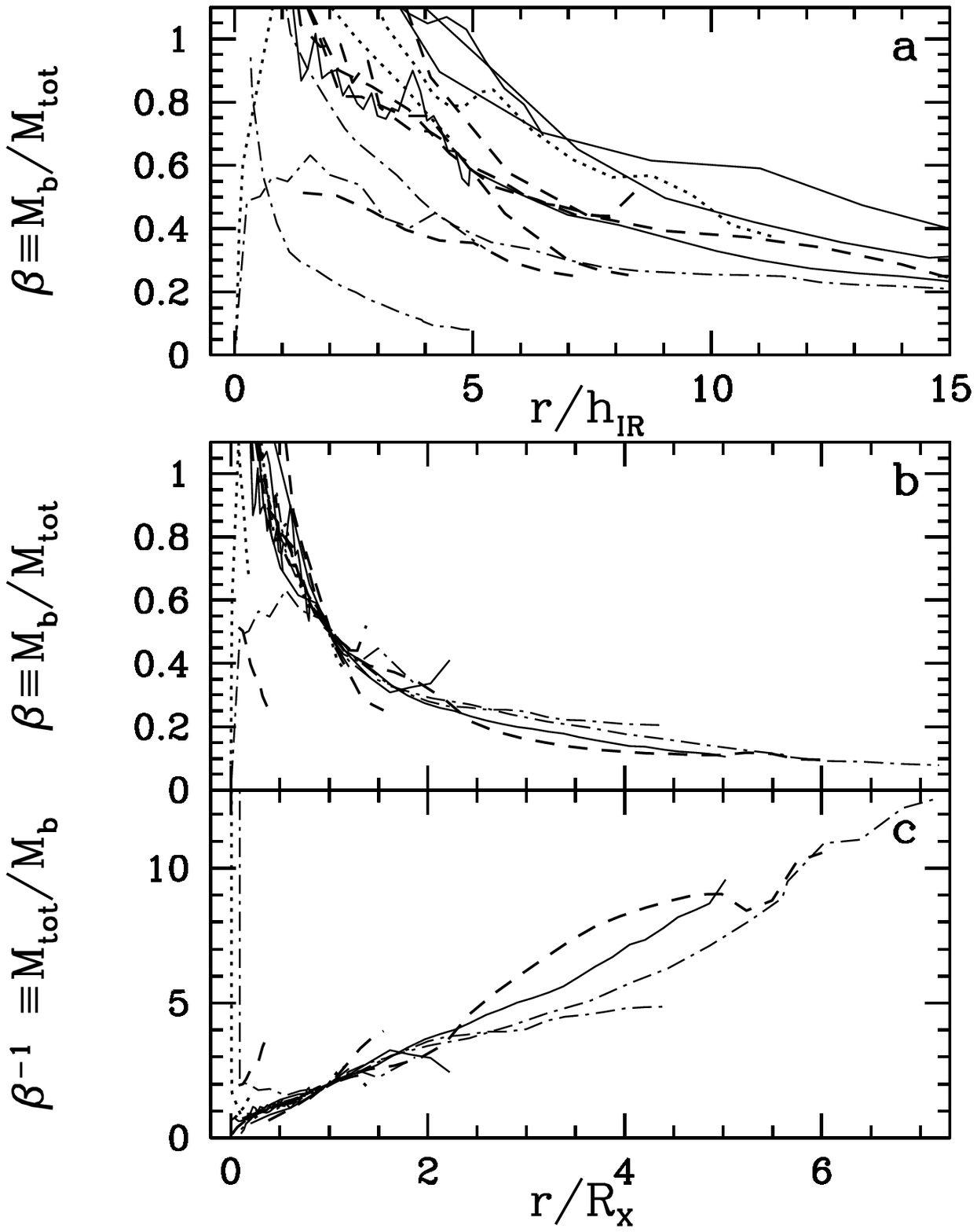}
\figcaption{{\small $\beta(r) \equiv M_b(r)/M_{tot}(r)$ and its inverse for
galaxies with an appreciable dark matter contribution.  
$\beta(r)$ is plotted versus $r/h_{IR}$ and 
$r/R_X$ in parts a and b, respectively; its inverse 
is plotted versus $r/R_X$ in panel c.  Different
line types correspond to the $V_{b,max}$ of the galaxies: $V_{b,max}>250$ km s$^{-1}$ (solid
line), $201<V_{b,max}\le250$ (dotted line), $120<V_{b,max}\le201$ (dashed line),
and $V_{b,max}\le120$ (dot-dash line line).}  
\label{fig:betaplot}}
\end{figure}

In Figure~\ref{fig:betaplot}, the dimensionless parameter
$\beta(r) \equiv M_b(r)/M_{tot}(r)$, where $M_b$ is the baryonic mass, 
and its inverse are plotted
for galaxies with an appreciable dark matter contribution.
It measures the fractional contribution of baryons to
the gravitational potential as a function of radius in a galaxy, and
is akin to the $\beta$ parameter
defined by \citet{salu} and similar parameters used in many other papers,
except here it is evaluated at all radii.  Where $\beta(r)=1$, the
baryonic mass of a galaxy accounts for its observed rotation
curve, $\beta(r)= 0.5$ at $r=R_X$, and
$\beta(r)=0$ where the dark matter accounts for its
observed rotation curve.  If the baryonic model over-predicts
the observed rotation curve, then $\beta(r) > 1$.
In Figure~\ref{fig:betaplot}a, $\beta(r)$ is plotted versus radius in
terms of $h_{IR}$.  For many galaxies, the observed rotation curves are
entirely accounted for by baryons in the inner
parts.  Beyond this region, baryonic mass falls off as dark matter
begins to dominate.  Other galaxies are dark matter dominated
throughout.  Curves in Figure~\ref{fig:betaplot}a have different line types that
correspond to ranges of $V_{b,max}$.  For most of the galaxies, as $V_{b,max}$
increases, so does the proportion of baryonic to dark
matter at all radii such that the fastest rotators
are observed to be dominated by baryons until quite far out into their
disks.  However, there is clearly much scatter about this trend; 
this conclusion can also be inferred from Figure~\ref{fig:reltns_dm}f.

In Figure~\ref{fig:betaplot}b, $\beta(r)$ is plotted versus radius in units of $R_X$.  
This choice of radial coordinate causes the curves
to overlap at $r=R_X$ (where $\beta(r)=0.5$), and thus
allows for a better comparison of their radial behavior.
In Figure~\ref{fig:betaplot}c, we plot the somewhat less intuitive 
function $\beta^{-1}(r)$ since it increases linearly with $r/R_X$.
There is little variation in the behavior of the curves in Figures~\ref{fig:betaplot}b,c.
Such regularity can be explained in terms of quasi-exponential disks and 
flat rotation curves as follows:  The rotation
curve of a galaxy is nearly flat beyond $r \sim 2 h_{IR}$, or it is at least
a slow function of $r$ and varies less than $V_b(r)$ in this region.  
In addition, much of the baryonic mass of a galaxy is enclosed at $r \sim 2 h_{IR}$,
which causes the baryonic rotation curves to be roughly
Keplerian ($V_b(r) \propto 1/r$) beyond this radius.  Therefore,
for all galaxies, beyond $\sim 2 h_{IR}$, it should be the case that 
$\beta(r) \equiv V_b^2/V_{tot}^2 \propto 1/r$, which is what we observe.
All the $\beta(r)$ curves overlap nicely when plotted versus radius
normalized to $R_X$ since this radius is located beyond
$2 h_{IR}$, and is generally in the falling part of the baryonic rotation 
curves.  We parametrize the trend of  $\beta^{-1}$ with $r/R_X$ in a simple universal manner that
holds for all galaxies: $\beta^{-1} = 1.71 (r/R_X) + 0.021$.

This relation is consistent with the prediction of \citet{palu}
that either the contribution of dark matter within the optical
radius of galaxies is small or that the distribution of dark matter
is coupled to that of the luminous matter.
This relation is also similar in spirit to the parameterization
of the mass discrepancy-acceleration relation of \citet{mcg4}.
Since $\beta^{-1}$ is the mass discrepancy, and $a=V^2/r \propto 1/r$
for flat rotation curves, a natural mass discrepancy-acceleration relation
arises.

\section{Comparison With Theories of Halo Density Distributions}
We compare the derived dark matter profiles
with an analytical function designed to parametrize the density profile
of dark matter halos in N-body simulations.
In particular, we compare our data to
the NFW formulation for dark matter halo density profiles \citep{nnfw} with and without taking into account 
adiabatic contraction of the halos.
To do this, for each galaxy, we fit its observed rotation curve
with a total mass rotation curve created from the addition of 
its baryonic rotation curve to a grid of halo models,
with a reduced $\chi^2$ statistic.  Uncertainties in the observed 
rotation curves are taken to be $10$ km s$^{-1}$; results do not differ
significantly if the error bars plotted in Figure 3 are used.
To test the most common implementation of dark matter contraction \citep{blum}, we 
adiabatically contract the grid of halo models according to the radial density distribution
of baryons in the galaxies following the formalism of \citet{dutt},
and perform the fits again.  The NFW fitting formula has 2 free parameters that we fit
for using a grid covering: $0.5 \le c_{200} \le 20$ and $0.5\,V_{tot,max} \le V_{200} \le 2.5\,V_{tot,max}$.

In Figure 17, we plot the best-fit halo models
for 8 example galaxies, and in Table 3 we list the best-fit parameters along with the
reduced $\chi^2$ of all the fits.  
Most of the fits are very poor; the average reduced $\chi^2$ values with and
without adiabatic contraction are: 7.1 and 4.2 for the
original color-$M/L$ relations, 14.0 and 7.3 for the $+0.1$ dex
renormalization of the color-$M/L$ relations, 4.8 and 3.4 for $-0.1$ dex, and 3.9 and 4.3
for $-0.3$ dex.  We do not perform fits for galaxies
when most of the observed rotation curve is accounted
for by the baryonic rotation curve.
\begin{figure*}
\figurenum{17}
\includegraphics[scale=0.8]{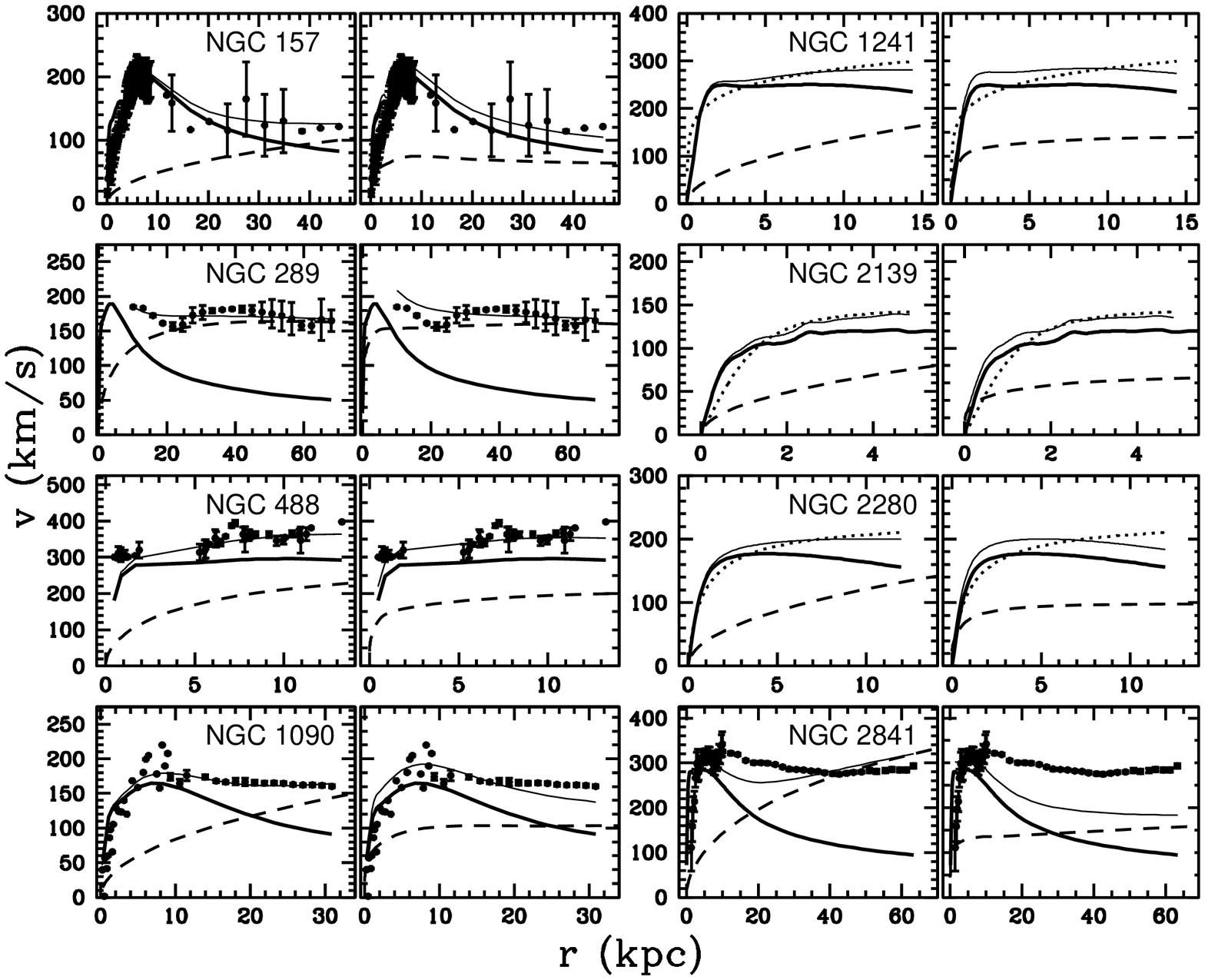}
\caption{{\small NFW halo fits for 8 example galaxies 
without (left) and with (right) adiabatic contraction.
Best-fit NFW models are plotted
as dashed lines, baryonic rotation curves as thick solid lines,
total rotation curves (sums of the best-fit NFW models
and the baryonic rotation curves) as thin solid lines,
and observed rotation curves as points. 
For those galaxies with a rotation curve
from Mathewson et al.\ 1992 that is modeled by Courteau 1997, we only plot the model
as a dotted line for clarity.}
\label{fig:halo_fits}}
\end{figure*}
\begin{figure}[!t]
\figurenum{18}
\hspace{-0.6in}
\includegraphics[height=4.2in, width=4.2in]{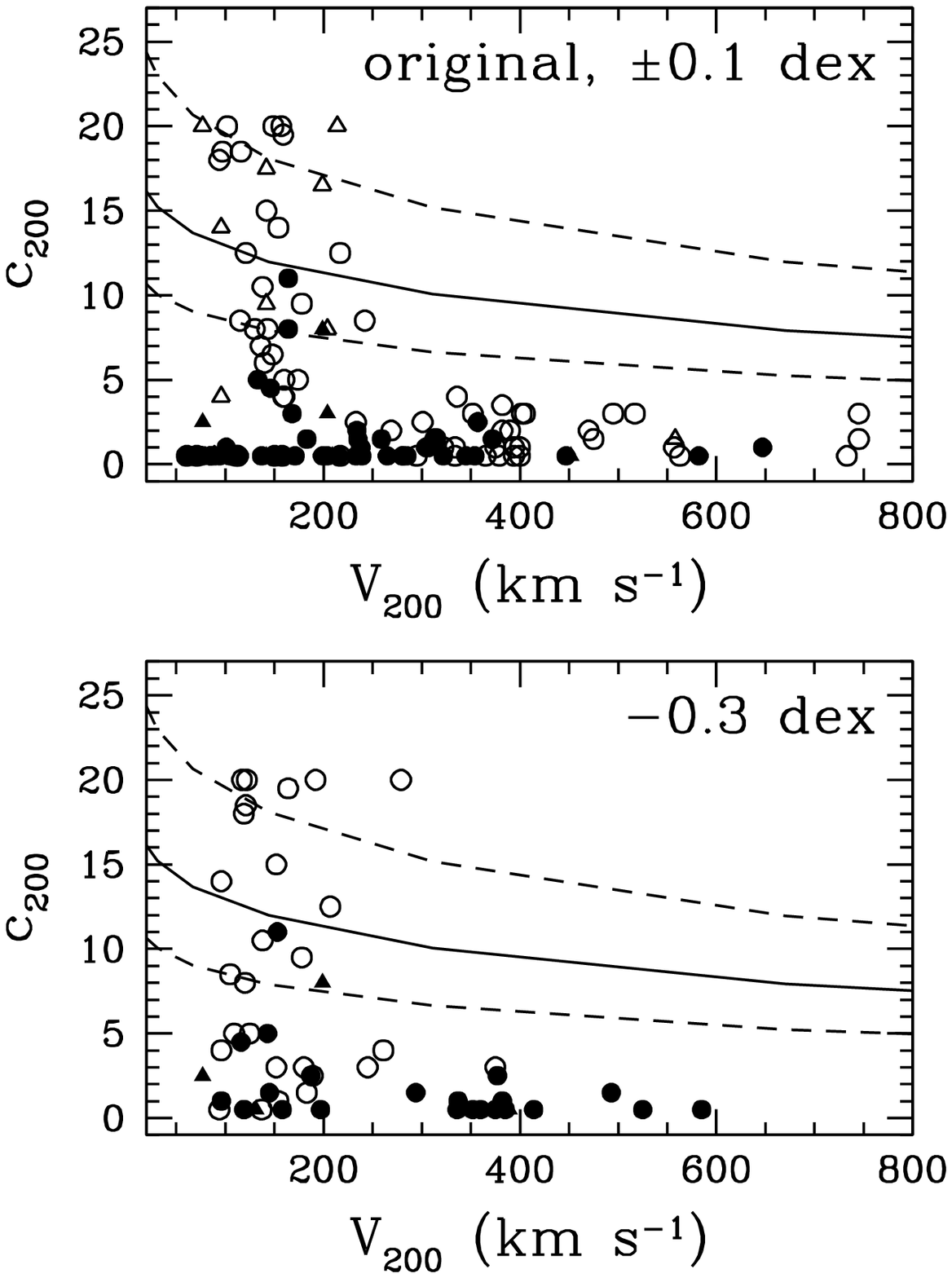}
\vspace{-0.5in}
\caption{{\small Best-fit parameters for NFW
fits with and without adiabatic contraction (filled and open symbols, 
respectively).  The median and the 68 per cent of the $c_{200}$ values
measured in numerical N-body simulations of dark matter halos are plotted
as solid and dashed lines, respectively. When the best-fit halo is at the edge 
of the parameter space searched, it is plotted as a triangle, otherwise it
is plotted as a circle.
In part a, best-fit parameters for the original and $\pm 0.1$ dex renormalizations 
of the color-$M/L$ relations are plotted; in part b, best-fit parameters for 
the -0.3 dex renormalization are plotted.} 
\label{fig:concenfits}}
\end{figure}

For baryonic rotation curves derived from the original color-$M/L$ relations,
nearly all fits to the NFW models without 
contraction have a smaller reduced $\chi^2$ than those where contraction was performed.
The exceptions are NGC\,4062 and NGC\,3992, which likely have submaximal disks.
For galaxies that have a large baryonic contribution 
to the inner parts of their rotation curves, adiabatic 
contraction over-contracts the inner parts of their dark matter halos 
such that the resulting total mass
rotation curves over-predict the measured rotation curves, as in \citet{wein}, for example.
For fits to the NFW models where the baryonic rotation curves were derived
from color-$M/L$ relations that were
renormalized by +0.1 dex, the situation is exactly the same. 
For -0.1 dex, 7 galaxies have better fits when contraction is taken into account,
and for -0.3 dex, 15 galaxies have a better fit.
This implementation of dark matter contraction works best when baryons do not account
for all of the inner parts of the observed rotation curve.

In Figure~\ref{fig:concenfits}, we plot the best-fit values of $V_{200}$ and $c_{200}$
and the range of these parameters found in N-body simulations (\citealt{bu01}; \citealt{eken}).
Even under the assumption of no adiabatic contraction, the derived concentrations
for these halos are low compared to those measured from simulations.
The concentrations for $-0.3$ dex are in slightly better agreement with
simulations, but are still quite low. In concert with this, \citet{mcg3} and \citet{mcg4}
found concentrations for NFW halos that are too low for a standard
$\Lambda$CDM universe.  In addition, \citet{alam} examined low surface brightness
galaxies and found their dark matter halos to be under-concentrated 
compared to what is expected from a standard $\Lambda$CDM universe, even if they assumed
the galaxies to be dark matter-dominated.

\section{Summary}
We have decomposed the rotation curves of 34 nearby bright spiral galaxies into baryonic
and dark matter components by applying color-$M/L$ relations to near-infrared
and optical photometry, and find the following:

\begin{itemize}

\item The dark-to-luminous matter distributions are self-similar
once scaled by $R_X$, the radius where the baryonic and dark matter contributions
to the rotation curve are equivalent.  This behavior is parameterized
by a simple function whose form is due to the quasi-exponential
nature of galaxy disks and rotation curves that are nearly flat after an initial rise.
This result is independent of the normalization of the color-$M/L$ relations.

\item The radii $R_{10}$ and $R_X$, where dark matter contributes $10 \%$ and
$50\%$ to the rotation of galaxies, respectively, correlate with galaxy properties.
The strongest correlation with $R_X$ is with the
maximum baryonic rotation speed such that galaxies with $R_X$ measurements that
lie further out in their disks rotate faster.  The next strongest correlations are
equivalently baryonic mass, observed rotation speed, and Hubble T-type.
Contrary to what is expected from previous studies \citep{debl,zava}, 
$B$-band central surface brightness is not found to be the main driving
force in relations with $R_X$ for this sample of bright galaxies.
The radii $R_X$ and $R_{10}$ move out in galaxy disks in tandem,
consistent with a self-similar dark-to-luminous matter distribution among galaxies.  

\item We confirm the normalization of the color-$M/L$ relations given in
\citet{bell}, which is analogous to an upper limit on the IMF based on the
dynamics of disk galaxies.  A more careful analysis of the data
used in both \citet{bdej} and \citet{bell} is performed, and 
data from this paper are included in the sample.

\item All but 4 of the 34 galaxies in our sample are close to maximal disk.  Two of these
4 galaxies have maximal disks within uncertainties.  
A prime example of a galaxy that cannot have a submaximal disk given the current
formulations of dark matter halos is NGC\,157 which has a pronounced hump-like
structure in its observed and baryonic rotation curves.

\item Maximum rotation velocities
predicted from the baryon distributions of galaxies ($V_{b,max}$) 
are tightly correlated with the observed maximum
rotation speeds ($V_{tot,max}$).  
Using this, a baryonic Tully-Fisher relation can be created based on two-passband surface photometry 
and a redshift alone (e.g., with the SDSS and color-$M/L$ relations for $g-r$ and $(M/L_r)_*$, or for
more distant redshift surveys with Hubble Space Telescope imaging).
Such a relation could possibly be calibrated for lower luminosity galaxies.


\item We find generally poor fits for the NFW parameterization for 
dark matter halos due to the significant baryonic contributions found in the
inner parts of most galaxies.  The concentrations of the best-fit NFW halos are therefore much 
lower than what is expected for galaxies in a standard $\Lambda$CDM 
universe.  This is even the case when the color-$M/L$ relations are renormalized by -0.3 dex.
In order to have better fits, a normalization even lower than -0.3 dex, 
where baryons contribute very little to the total mass in the inner parts 
of rotation curves, would have to be implemented.   
  Adiabatic contraction, as it is normally implemented,
 makes these fits worse by adding more dark matter to the inner parts of galaxies.

\end{itemize}

\acknowledgements
We would like to thank Richard Pogge, James Pizagno, 
James Bullock, Jay Frogel, Chris Kochanek, and David Weinberg for valuable discussions.  
The referee, Stacy McGaugh, is  thanked for
constructive comments that greatly improved the quality of this paper.
We are grateful to the following authors who provided rotation curves in
tabular form via email: Kor Begeman, Gianfranco Gentile, Thilo Kranz,
Povilas Palunas, Stuart Ryder, Michele Thornley, and Wilfred Walsh. 
We are grateful to Marc Verheijen who made available his 
high-quality imaging and rotation curves of galaxies in the Ursa Major cluster.
SAK would like to acknowledge financial support from The Space
Telescope Science Institute Director's Discretionary Research Fund (DDRF).
We thank the CTIO TAC for generous allocation of time for The Ohio State University 
Galaxy Survey and the many people over the years who helped collect 
these observations.  Funding for The Ohio State University Bright Spiral Galaxy Survey was 
provided by grants from The National Science Foundation 
(grants AST-9217716 and AST-9617006), with additional funding by The 
Ohio State University.  

This paper makes use of data from both the Sloan Digital Sky Survey and the
Two Micron All Sky Survey.  The Two Micron All Sky Survey is a joint 
project of the University of Massachusetts and the Infrared Processing 
and Analysis 
Center/California Institute of Technology, funded by the National 
Aeronautics and Space Administration and the National Science Foundation.
Funding for the creation and distribution of
the SDSS Archive has been provided by the Alfred P. Sloan Foundation,
the Participating Institutions, the National Aeronautics and Space
Administration, the National Science Foundation, the U.S. Department
of Energy, the Japanese Monbukagakusho, and the Max Planck
Society. The SDSS Web site is http://www.sdss.org/.
The SDSS is managed by the Astrophysical Research Consortium (ARC) for
the Participating Institutions. The Participating Institutions are The
University of Chicago, Fermilab, the Institute for Advanced Study, the
Japan Participation Group, The Johns Hopkins University, Los Alamos
National Laboratory, the Max-Planck-Institute for Astronomy (MPIA),
the Max-Planck-Institute for Astrophysics (MPA), New Mexico State
University, University of Pittsburgh, Princeton University, the United
States Naval Observatory, and the University of Washington.

This research has made use of NASA's Astrophysics Data System, 
the NASA/IPAC Extragalactic Database (NED), and the HyperLeda database 
and the VizieR catalog access tool. NED that is operated by the Jet Propulsion 
Laboratory, California Institute of Technology, 
under contract with the National Aeronautics and Space Administration.
This research also made use of the HyperLeda database and the VizieR 
catalog access tool, CDS, Strasbourg, France.

\begin{deluxetable}{ccccclll}
\tabletypesize{\small}
\tablehead{
\colhead{} &
\colhead{} &
\colhead{} &
\colhead{} &
\colhead{} &
\colhead{Galaxy} &
\colhead{Tracer\tablenotemark{a}} &
\colhead{Reference(s)}
}
\startdata
&&&&&\object{NGC 157}  &FP H$\alpha$, \ion{H}{1} &\citealt{frid}, \citealt{ryd2}\tablenotemark{b}\\
&&&&&\object{NGC 289}  &\ion{H}{1}  &\citealt{wals}\tablenotemark{b}\\
&&&&&\object{NGC 488}  &H$\alpha$ &\citealt{pet3}\tablenotemark{b,c}\\
&&&&&\object{NGC 908}  &H$\alpha$ &\citealt{math}\\
&&&&&\object{NGC 1087} &H$\alpha$ &\citealt{rub1}\\
&&&&&\object{NGC 1090} &H$\alpha$, \ion{H}{1}   &\citealt{cour}, \citealt{gent}\\
&&&&&\object{NGC 1241} &H$\alpha$ &\citealt{math}\\
&&&&&\object{NGC 1385} &H$\alpha$ &\citealt{math}\\
&&&&&\object{NGC 1559} &H$\alpha$ &\citealt{math}\\
&&&&&\object{NGC 1832} &H$\alpha$ &\citealt{math}\\
&&&&&\object{NGC 2090} &H$\alpha$ &\citealt{math}\\
&&&&&\object{NGC 2139} &H$\alpha$ &\citealt{math}\\
&&&&&\object{NGC 2280} &H$\alpha$ &\citealt{math}\\
&&&&&\object{NGC 2841} &FP H$\alpha$, \ion{H}{1} &\citealt{blai}, \citealt{gira}\tablenotemark{c}\\
&&&&&\object{NGC 3198} &FP H$\alpha$, \ion{H}{1} &\citealt{corr}\tablenotemark{b}, \citealt{vana}\\
&&&&&\object{NGC 3223} &H$\alpha$  &\citealt{math}\\
&&&&&\object{NGC 3319} &\ion{H}{1} &\citealt{moor}\tablenotemark{b,c}\\
&&&&&\object{NGC 3521} &\ion{H}{1}  &\citealt{sand}\tablenotemark{c}\\
&&&&&\object{NGC 3726} &\ion{H}{1}  &\citealt{verh}\tablenotemark{b}\\
&&&&&\object{NGC 3893} &H$\alpha$, \ion{H}{1} &\citealt{kran}, \citealt{verh}\tablenotemark{b}\\
&&&&&\object{NGC 3949} &\ion{H}{1} &\citealt{verh}\tablenotemark{b}\\
&&&&&\object{NGC 3953} &\ion{H}{1}  &\citealt{verh}\tablenotemark{b}\\
&&&&&\object{NGC 3992} &\ion{H}{1} &\citealt{verh}\tablenotemark{b}\\
&&&&&\object{NGC 4051} &\ion{H}{1} &\citealt{verh}\tablenotemark{b}\\
&&&&&\object{NGC 4062} &H$\alpha$ &\citealt{rub1}\\
&&&&&\object{NGC 4138} &\ion{N}{2}, \ion{H}{1}  &\citealt{jore}\tablenotemark{c},\citealt{verh}\tablenotemark{b}\\
&&&&&\object{NGC 4651} &H$\alpha$ &\citealt{rub9}\\
&&&&&\object{NGC 4698} &H$\alpha$  &\citealt{rub9}\\
&&&&&\object{NGC 5371} &\ion{H}{1} &\citealt{bege}\\
&&&&&\object{NGC 5806} &H$\alpha$ &\citealt{cour}\\
&&&&&\object{NGC 6300} &FP H$\alpha$, \ion{H}{1} &\citealt{buta}, \citealt{ryd1}\\
&&&&&\object{NGC 7083} &H$\alpha$ &\citealt{math}\\
&&&&&\object{NGC 7217} &H$\alpha$ &\citealt{rub1}\\
&&&&&\object{NGC 7606} &H$\alpha$ &\citealt{math}\\
\enddata
\tablenotetext{a}{The notation ``FP H$\alpha$'' is used for rotation
curves derived from Fabry-Perot measurements of H$\alpha$.}
\tablenotetext{b}{Errors are taken as
the difference in velocity between the approaching and receding sides.}
\tablenotetext{c}{The rotation curve has been extracted
  electronically from a plot in the referenced paper.}
\end{deluxetable}

\begin{deluxetable}{cccccccccc}
\tabletypesize{\small}
\tablehead{
\colhead{}
&\colhead{}
&\colhead{}
&\colhead{}
&\colhead{Observed}
&\multicolumn{3}{c}{Baryonic Matter}
&\multicolumn{2}{c}{Dark Matter}\\
\cline{5-5}
\cline{6-8}
\cline{9-10}

\colhead{}
&\colhead{}
&\colhead{}
&\colhead{}
&\colhead{$V_{tot,max}$}
&\colhead{$V_{b,max}$}
&\colhead{$R(V_{b,max})$}
&\colhead{$M_b$}
&\colhead{$R_{10}$}
&\colhead{$R_{\rm X}$}\\

\colhead{}
&\colhead{}
&\colhead{}
&\colhead{}
&\colhead{}
&\colhead{$\Delta_{+0.1 dex}$}
&\colhead{}
&\colhead{$\Delta_{+0.1 dex}$}
&\colhead{$\Delta_{+0.1 dex}$}
&\colhead{$\Delta_{+0.1 dex}$}\\

\colhead{}
&\colhead{}
&\colhead{}
&\colhead{}
&\colhead{}
&\colhead{$\Delta_{-0.1 dex}$}
&\colhead{}
&\colhead{$\Delta_{-0.1 dex}$}
&\colhead{$\Delta_{-0.1 dex}$}
&\colhead{$\Delta_{-0.1 dex}$}\\

\colhead{}
&\colhead{}
&\colhead{}
&\colhead{}
&\colhead{}
&\colhead{$\Delta_{-0.3 dex}$}
&\colhead{}
&\colhead{$\Delta_{-0.3 dex}$}
&\colhead{$\Delta_{-0.3 dex}$}
&\colhead{$\Delta_{-0.3 dex}$}\\

\colhead{}
&\colhead{}
&\colhead{}
&\colhead{Galaxy}
&\colhead{(km/s)}
&\colhead{(km/s)}
&\colhead{(kpc)}
&\colhead{($10 ^{10} M_{\odot}$)}
&\colhead{(kpc)}
&\colhead{(kpc)}
}
\startdata
&&&NGC 157 &205 &205 &  6.2 &   7.1 &\nodata & \nodata \\
&&& &  & +25 &        &   +1.8 &\nodata &\nodata\\ 
&&& &  & -22 &        &   -1.4 &\nodata &\nodata \\
&&& &  & -60 &        &   -3.5 &\nodata &\nodata \\
&&&NGC 289 &182 &189 &  3.6 &   4.1 &  9.1 & 11.3 \\
&&& &  & +23 &        &   +1.0 & +1.7  &  +4.2 \\
&&& &  & -20 &        &   -0.8 &  -1.8 &  -2.7 \\
&&& &  & -55 &        &   -2.0 &  -4.0 &  -4.7 \\
&&&NGC 488 &350 &296 & 10.1 &   31.1 &\nodata & \nodata\\ 
&&& &  & +36 &        &   +7.4 &\nodata &\nodata \\
&&& &  & -32 &        &   -5.9 &\nodata &\nodata \\
&&& &  & -86 &        &   -14.3 &\nodata &\nodata \\
&&&NGC 908 &200 &201 &  7.3 &   7.2 &\nodata & \nodata \\
&&& &  & +24 &        &   +1.8 &\nodata &\nodata \\
&&& &  & -22 &        &   -1.5 &\nodata &\nodata \\
&&& &  & -59 &        &   -3.5 &\nodata &\nodata \\
&&&NGC 1087 &136 &141 &  5.9 &   2.8 &\nodata & \nodata \\
&&& &  & +17 &        &   +0.7 &\nodata &\nodata \\
&&& &  & -15 &        &   -0.6 &\nodata &\nodata \\
&&& &  & -41 &        &   -1.4 &\nodata &\nodata \\
&&&NGC 1090 &170 &165 &  7.3 &   5.4 & 16.4 & 23.2 \\
&&& &  & +20 &        &   +1.4 &  +4.4 &  +7.8\tablenotemark{c} \\
&&& &  & -18 &        &   -1.1 &  -5.3 &  -4.9 \\
&&& &  & -49 &        &   -2.7 & -12.5 & -18.2 \\
&&&NGC 1241 &300 &250 &  7.9 &   19.2 & 12.4 & \nodata \\
&&& &  & +30 &        &   +4.8 &  +4.4\tablenotemark{d} &\nodata \\
&&& &  & -27 &        &   -3.8 &  -5.6 &\nodata \\
&&& &  & -73 &        &   -9.2 &  -12.4\tablenotemark{d} &\nodata \\
&&&NGC 1385 &140 &133 &  4.0 &   2.6 &\nodata & \nodata \\
&&& &  & +16 &        &   +0.6 &\nodata &\nodata \\
&&& &  & -14 &        &   -0.5 &\nodata &\nodata \\
&&& &  & -39 &        &   -1.2 &\nodata &\nodata \\
&&&NGC 1559 &150 &136 &  4.5 &   2.3 &\nodata & \nodata \\
&&& &  & +17 &        &   +0.6 &\nodata &\nodata \\
&&& &  & -15 &        &   -0.5 &\nodata &\nodata \\
&&& &  & -40 &        &   -1.1 &\nodata &\nodata \\
&&&NGC 1832 &200 &209 &  3.8 &   5.4 &\nodata & \nodata\\ 
&&& &  & +25 &        &   +1.4 &\nodata &\nodata \\
&&& &  & -23 &        &   -1.1 &\nodata &\nodata \\
&&& &  & -61 &        &   -2.7 &\nodata &\nodata \\
&&&NGC 2090\tablenotemark{a} &160 &153 &  1.6 &   1.6 &\nodata & \nodata \\
&&& &  & +19 &        &   +0.4 &\nodata &\nodata \\
&&& &  & -16 &        &   -0.3 &\nodata &\nodata \\
&&& &  & -45 &        &   -0.8 &\nodata &\nodata \\
&&&NGC 2139 &140 &121 &  4.5 &   1.9 &\nodata & \nodata\\ 
&&& &  & +15 &        &   +0.5 &\nodata &\nodata \\
&&& &  & -13 &        &   -0.4 &\nodata &\nodata \\
&&& &  & -35 &        &   -0.9 &\nodata &\nodata \\
&&&NGC 2280\tablenotemark{a} &210 &177 &  6.9 &   6.8 &  8.7 & \nodata \\
&&& &  &  +5 &        &   +1.8 &  +1.8 &\nodata \\
&&& &  & -19 &        &   -1.4 &  -3.5 &\nodata \\
&&& &  & -32 &        &   -3.4 &  -5.4 &\nodata \\
&&&NGC 2841 &325 &284 &  4.7 &   13.0 &  9.0 & 12.6 \\
&&& &  & +49 &        &   +3.3 &  +3.4 &  +5.8 \\
&&& &  & -20 &        &   -2.6 &  -0.6 &  -2.9 \\
&&& &  & -35 &        &   -6.4 &  -6.2 &  -3.7 \\
&&&NGC 3198 &152 &120 &  5.6 &   2.3 &  6.8 & 10.5 \\
&&& &  & +14 &        &   +0.6 &  +3.1 &  +2.6 \\
&&& &  & -13 &        &   -0.5 &  -2.1 &  -2.5 \\
&&& &  & -35 &        &   -1.1 & -4.6 &  -7.3 \\
&&&NGC 3223 &320 &314 & 10.4 &   30.8 &\nodata & \nodata \\
&&& &  & +38 &        &   +7.6 &\nodata &\nodata \\
&&& &  & -34 &        &   -6.1 &\nodata &\nodata \\
&&& &  & -92 &        &   -14.7 &\nodata &\nodata \\
&&&NGC 3319\tablenotemark{b} &132 & 50 &  9 &  0.7 &  2.6 &  3.5 \\
&&& &  &  +7 &        &   +0.1 &  +0.7 &  +1.1 \\
&&& &  &  -5 &        &   -0.1 &  -0.6 &  -0.8 \\
&&& &  & -14 &        &   -0.3 &  -2.6\tablenotemark{d} & -3.5\tablenotemark{d} \\
&&&NGC 3521 &221 &263 &  3.1 &   8.0 & 10.3 & 13.3 \\
&&& &  & +32 &        &   +1.9 &  +2.3 &  +4.2 \\
&&& &  & -29 &        &   -1.5 &  -1.8 &  -2.4 \\
&&& &  & -34 &        &   -3.6 &  -5.8\tablenotemark{d} &  -6.3 \\
&&&NGC 3726 &169 &144 & 10.8 &   6.0 & 20.4 & 25.3 \\
&&& &  & +22 &        &   +1.5 &  +3.2 &  +3.5 \\
&&& &  & -12 &        &   -1.2 & -10.6 &  -2.2 \\
&&& &  & -42 &        &   -2.9 & -20.4\tablenotemark{d} & -16.3 \\
&&&NGC 3893 &210 &186 &  5.3 &   6.8 & 13.3 & 19.4 \\
&&& &  & +23 &        &   +1.6 &  +4.5 &  +2.6\tablenotemark{c} \\
&&& &  & -20 &        &   -1.3 &  -5.5 &  -4.7 \\
&&& &  & -54 &        &   -3.2 & -10.3 & -13.4 \\
&&&NGC 3949 &169 &158 &  3.9 &   2.7 &  7.4 & \nodata \\
&&& &  & +18 &        &   +0.6 &  +0.7\tablenotemark{d} &\nodata \\
&&& &  & -17 &        &   -0.5 &  -1.6 &\nodata \\
&&& &  & -46 &        &   -1.2 &  -7.4\tablenotemark{d} &\nodata \\
&&&NGC 3953 &225 &227 &  7.7 &   12.0 & 17.8 & \nodata \\
&&& &  & +28 &        &   +3.1 &  +0.3\tablenotemark{c} &\nodata \\
&&& &  & -25 &        &   -2.4 &  -4.2 &\nodata \\
&&& &  & -66 &        &   -5.9 & -14.0 &\nodata \\
&&&NGC 3992 &272 &188 & 13.1 &   11.2 &\nodata &\nodata \\
&&& &  & +21 &        &   +2.8 &\nodata &\nodata \\
&&& &  & -21 &        &   -2.3 &\nodata &\nodata \\
&&& &  & -56 &        &   -5.5 &\nodata &\nodata \\
&&&NGC 4051 &170 &167 &  0.3 &   2.4 &  5.8 & \nodata \\
&&& &  & +20 &        &   +0.6 & +2.1 &\nodata \\
&&& &  & -18 &        &   -0.5 &  -1.3 &\nodata \\
&&& &  & -49 &        &   -1.1 & -5.8\tablenotemark{d} &\nodata \\
&&&NGC 4062 &162 &110 &  3.4 &   1.1 &  0\tablenotemark{d} &  3.9 \\
&&& &  & +14 &        &   +0.3 &  +1 &  +2.4 \\
&&& &  & -11 &        &   -0.2 &  0\tablenotemark{d} &  -2.0 \\
&&& &  & -32 &        &   -0.6 & 0\tablenotemark{d} & -3.9\tablenotemark{d} \\
&&&NGC 4138 &195 &272 & 1.0 &   4.2 &  7.5 & 17.8 \\
&&& &  & +33 &        &   +1.0 &  +8.9 &  +3.1 \\
&&& &  & -30 &        &   -0.8 &  -1.4 &  -5.8 \\
&&& &  & -79 &        &   -2.0 & -7.5\tablenotemark{d} & -12.6 \\
&&&NGC 4651 &210 &173 & 2.6  &   3.7 &\nodata & \nodata \\
&&& &  & +32 &        &   +0.9 &\nodata &\nodata \\
&&& &  & -10 &        &   -0.8 &\nodata &\nodata \\
&&& &  & -45 &        &   -1.8 &\nodata &\nodata \\
&&&NGC 4698 &220 &223 &  3.8 &   6.5 &\nodata & \nodata \\
&&& &  & +14 &        &   +1.6 &\nodata &\nodata \\
&&& &  & -35 &        &   -1.3 &\nodata &\nodata \\
&&& &  & -68 &        &   -3.2 &\nodata &\nodata \\
&&&NGC 5371 &242 &289 & 13.9 &   33.5 & \nodata & \nodata \\
&&& &  & +21 &        &   +8.5 & \nodata &\nodata \\
&&& &  & -43 &        &   -6.7 & \nodata &\nodata \\
&&& &  & -87 &        &   -16.4 & \nodata &\nodata \\
&&&NGC 5806 &200 &190 &  1.1 &   5.3 &\nodata & \nodata \\
&&& &  & +27 &        &   +1.3 &\nodata &\nodata \\
&&& &  & -17 &        &   -1.1 &\nodata &\nodata \\
&&& &  & -27 &        &   -2.6 &\nodata &\nodata \\
&&&NGC 6300 &208 &220 &  4.3 &   6.7 & 13.7 & 20.4 \\
&&& &  & +12 &        &   +1.7 &  +5.8 &  +2.6 \\
&&& &  & -35 &        &   -1.4 &  -1.3 &  -5.5 \\
&&& &  & -67 &        &   -3.3 &  -8.8 & -13.5 \\
&&&NGC 7083 &210 &245 &  6.5 &   13.1 &\nodata & \nodata \\
&&& &  & +30 &        &   +3.3 &\nodata &\nodata \\
&&& &  & -27 &        &   -2.7 &\nodata &\nodata \\
&&& &  & -72 &        &   -6.4 &\nodata &\nodata \\
&&&NGC 7217 &284 &283 &  2.0 &   7.9 &\nodata & \nodata \\
&&& &  & +35 &        &   +2.0 &\nodata &\nodata \\
&&& &  & -31 &        &   -1.6 &\nodata &\nodata \\
&&& &  & -84 &        &   -3.9 &\nodata &\nodata \\
&&&NGC 7606 &280 &228 & 9.8 &   12.9 &  2.2 & \nodata \\
&&& &  & +28 &        &   +3.2 &  +0.8\tablenotemark{c} &\nodata \\
&&& &  & -25 &        &   -2.6 &  -0.7 &\nodata \\
&&& &  & -67 &        &   -6.2 &  -2.2\tablenotemark{d} &\nodata \\
\enddata
\tablenotetext{a}{Quantities are calculated from $B-V$ instead of $B-R$}
\tablenotetext{b}{Quantities are calculated from $(M/L)_{*,H}$ instead of $(M/L)_{*,K}$}
\tablenotetext{c}{Upper limit}
\tablenotetext{d}{\,Lower limit}
\end{deluxetable}

\begin{deluxetable}{cccccccccccccc}
\tabletypesize{\small}
\tablehead{

\colhead{}
&\multicolumn{3}{c}{origonal color-$M/L$}
&\multicolumn{3}{c}{+0.1 dex}
&\multicolumn{3}{c}{-0.1 dex}
&\multicolumn{3}{c}{-0.3 dex}\\

\colhead{}
&\colhead{$c$}
&\colhead{$V_{200}$}
&\colhead{$\chi^2$}
&\colhead{$c$}
&\colhead{$V_{200}$}
&\colhead{$\chi^2$}
&\colhead{$c$}
&\colhead{$V_{200}$}
&\colhead{$\chi^2$}
&\colhead{$c$}
&\colhead{$V_{200}$}
&\colhead{$\chi^2$}\\

\colhead{Galaxy}
&\colhead{}
&\colhead{(km/s)}
&\colhead{}
&\colhead{}
&\colhead{(km/s)}
&\colhead{}
&\colhead{}
&\colhead{(km/s)}
&\colhead{}
&\colhead{}
&\colhead{(km/s)}
&\colhead{}\\
}
\startdata
NGC 157                   &0.5     &111                   &9.5  &0.5     &61\tablenotemark{a}   &19.9 &4.0     &96\tablenotemark{a}   &6.7  &16.0    &96                    &5.8\\ 
                          &0.5     &61\tablenotemark{a}   &13.8 &0.5     &61\tablenotemark{a}   &28.2 &0.5     &61\tablenotemark{a}   &8.2  &2.0     &131                   &7.3\\
NGC 289                   &8.0     &143                   &0.7  &6.5     &148                   &0.8  &10.5    &138                   &0.8  &11.5    &138                   &0.8\\
                          &3.0     &168                   &1.2  &1.0     &238                   &1.7  &5.0     &133                   &0.8  &7.5     &143                   &0.7\\
NGC 488                   &16.5    &199\tablenotemark{a}  &9.1  &8.0     &204                   &6.5  &20.0    &214\tablenotemark{a}  &12.6 &20.0    &279                   &22.7\\
                          &3.0     &204                   &5.3  &0.5     &199\tablenotemark{a}  &6.8  &8.0     &199\tablenotemark{a}  &5.2  &18.5    &199\tablenotemark{a}  &6.1\\
NGC 908                   &\nodata &\nodata           &\nodata  &\nodata &\nodata &\nodata            &\nodata &\nodata   &\nodata          &20.0    &116                   &8.7\\
                          &\nodata &\nodata           &\nodata  &\nodata &\nodata &\nodata            &\nodata &\nodata   &\nodata          &14.0    &96\tablenotemark{a}   &6.5\\
NGC 1087                  &\nodata &\nodata           &\nodata  &\nodata &\nodata &\nodata            &\nodata &\nodata   &\nodata          &16.5    &68\tablenotemark{a}   &1.3\\ 
                          &\nodata &\nodata           &\nodata  &\nodata &\nodata &\nodata            &\nodata &\nodata   &\nodata          &4.0     &68\tablenotemark{a}   &1.5\\
NGC 1090                  &1.0     &400                   &6.4  &0.5     &400                   &9.7  &5.0     &160                   &5.3  &12.0    &125                   &4.7\\
                          &0.5     &105                   &11.4 &0.5     &80\tablenotemark{a}   &15.6 &0.5     &215                   &8.7  &1.0     &360                   &6.6\\
NGC 1241                  &1.5     &745                   &4.7  &0.5     &150\tablenotemark{a}  &9.3  &3.0     &745                   &3.4  &20.0    &180                   &3.4\\ 
                          &0.5     &150\tablenotemark{a}  &7.3  &0.5     &150\tablenotemark{a}  &22.2 &0.5     &354                   &4.0  &1.5     &585                   &2.3\\
NGC 1385                  &\nodata &\nodata           &\nodata  &\nodata &\nodata &\nodata            &\nodata &\nodata   &\nodata          &4.0     &328                   &0.1\\
                          &\nodata &\nodata           &\nodata  &\nodata &\nodata &\nodata            &\nodata &\nodata   &\nodata          &0.5     &143                   &0.7\\  
NGC 1559                  &\nodata &\nodata           &\nodata  &\nodata &\nodata &\nodata            &\nodata &\nodata   &\nodata          &3.5     &350\tablenotemark{a}  &5.1\\
                          &\nodata &\nodata           &\nodata  &\nodata &\nodata &\nodata            &\nodata &\nodata   &\nodata          &0.5     &350\tablenotemark{a}  &7.0\\
NGC 1832                  &0.5     &89\tablenotemark{a}   &8.8  &\nodata &\nodata &\nodata            &0.5     &394                   &4.2  &20.0    &94                    &4.0\\
                          &0.5     &89\tablenotemark{a}   &19.1 &\nodata &\nodata &\nodata            &0.5     &394                   &6.9  &0.5     &389\tablenotemark{a}  &3.7\\  
NGC 2090                  &\nodata &\nodata           &\nodata  &\nodata &\nodata &\nodata            &20.0    &77\tablenotemark{a}   &0.8  &20.0    &122                   &1.2\\
                          &\nodata &\nodata           &\nodata  &\nodata &\nodata &\nodata            &2.5     &77\tablenotemark{a}   &0.4  &13.5    &77\tablenotemark{a}   &0.3\\
NGC 2139                  &2.5     &301                   &1.5  &0.5     &71\tablenotemark{a}   &2.2  &4.0     &336                   &1.1  &7.5     &261                   &0.8\\
                          &0.5     &71\tablenotemark{a}   &2.7  &0.5     &71\tablenotemark{a}   &5.6  &0.5     &156                   &2.5  &2.0     &336                   &2.3\\
NGC 2280                  &2.0     &470                   &1.2  &1.5     &475                   &0.5  &3.0     &495                   &0.2  &5.0     &375                   &0.2\\
                          &0.5     &105\tablenotemark{a}  &3.9  &0.5     &105\tablenotemark{a}  &2.4  &0.5     &265                   &2.7  &1.0     &525                   &2.6\\
NGC 2841                  &8.5     &242                   &13.4 &3.0     &402                   &27.6 &12.5    &217                   &11.0 &15.0    &207                   &7.0\\
                          &0.5     &582                   &23.3 &0.5     &452\tablenotemark{a}  &53.9 &1.0     &647                   &18.6 &3.5     &337                   &11.6\\
NGC 3198                  &6.0     &140                   &1.7  &4.0     &160                   &2.8  &8.0     &130                   &1.1  &11.5    &120                   &0.6\\
                          &0.5     &345                   &4.0  &0.5     &285                   &6.5  &1.5     &235                   &2.8  &5.0     &145                   &1.5\\ 
NGC 3223                  &0.5     &733                   &5.3  &0.5     &158\tablenotemark{a}  &14.8 &9.5     &178                   &5.0  &20.0    &178                   &7.1\\
                          &0.5     &158\tablenotemark{a}  &10.1 &0.5     &158\tablenotemark{a}  &39.9 &0.5     &233                   &3.7  &11.5    &158                   &2.6\\
NGC 3319                  &2.0     &269                   &0.3  &1.5     &314                   &0.3  &2.5     &233                   &0.2  &3.5     &189                   &0.2\\  
                          &1.0     &304                   &0.9  &1.5     &314                   &1.3  &1.5     &259                   &0.7  &1.5     &294                   &0.4\\  
NGC 3521                  &12.5    &121                   &3.1  &7.0     &136                   &4.3  &18.5    &116                   &2.6  &20.0    &121                   &2.2\\
                          &0.5     &281                   &4.1  &0.5     &171                   &6.2  &4.5     &146                   &3.5  &13.5    &116                   &2.1\\ 
NGC 3726                  &1.0     &334                   &1.0  &0.5     &379                   &3.6  &1.0     &394                   &0.3  &6.0     &154                   &0.6\\
                          &0.5     &159                   &5.9  &0.5     &94                    &10.2 &0.5     &239                   &2.9  &0.5     &414                   &0.4\\
NGC 3893                  &4.0     &159                   &2.3  &0.5     &334                   &4.4  &18.0    &94                    &2.2  &20.0    &119                   &4.0\\
                          &0.5     &74                    &3.7  &0.5     &74\tablenotemark{a}   &8.9  &0.5     &204                   &2.3  &8.5     &119                   &2.1\\
NGC 3949                  &2.0     &390                   &2.2  &0.5     &365                   &5.5  &3.0     &405                   &0.7  &7.0     &245                   &0.03\\
                          &0.5     &85\tablenotemark{a}   &4.5  &0.5     &85\tablenotemark{a}   &10.8 &0.5     &280                   &2.9  &2.0     &385                   &0.8\\
NGC 3953                  &0.5     &563\tablenotemark{a}  &2.0  &0.5     &113\tablenotemark{a}  &6.5  &1.5     &558                   &0.3  &9.5     &183                   &0.02\\
                          &0.5     &113\tablenotemark{a}  &5.3  &0.5     &113\tablenotemark{a}  &22.9 &1.5     &183                   &2.2  &1.0     &493                   &0.2\\
NGC 3992                  &20.0    &149                   &0.3  &14.0    &154                   &0.3  &19.5    &159                   &0.4  &20.0    &164                   &1.2\\
                          &8.0     &164                   &0.3  &2.0     &234                   &0.3  &11.0    &164                   &0.3  &20.0    &153                   &0.4\\ 
NGC 4051                  &2.0     &382                   &1.9  &1.0     &322                   &2.9  &3.0     &352                   &1.3  &10.0    &152                   &1.0\\
                          &0.5     &137                   &3.8  &0.5     &77\tablenotemark{a}   &4.9  &0.5     &322                   &2.5  &2.0     &352                   &1.7\\
NGC 4062                  &18.5    &97                    &1.7  &3.5     &382                   &1.5  &20.0    &102                   &1.7  &20.0    &117                   &2.0\\ 
                          &1.5     &372                   &0.6  &0.5     &352\tablenotemark{a}  &1.1  &2.5     &357                   &0.5  &4.0     &377                   &0.7\\
NGC 4138                  &1.0     &375\tablenotemark{a}  &15.7 &0.5     &295                   &30.1 &8.5     &115                   &9.9  &20.0    &105                   &7.3\\  
                          &0.5     &75\tablenotemark{a}   &23.6 &0.5     &75\tablenotemark{a}   &43.0 &0.5     &75\tablenotemark{a}   &15.4 &1.0     &375                   &10.8\\
NGC 4651                  &3.0     &517                   &2.3  &1.0     &557\tablenotemark{a}  &4.9  &15.0    &142                   &2.0  &20.0    &152                   &3.0\\
                          &0.5     &112\tablenotemark{a}  &6.1  &0.5     &112\tablenotemark{a}  &11.1 &1.0     &307                   &5.8  &3.5     &382                   &5.7\\
NGC 4698                  &\nodata &\nodata  &\nodata     &\nodata &\nodata &\nodata            &\nodata &\nodata   &\nodata                &3.5     &525                   &19.2\\ 
                          &\nodata &\nodata  &\nodata     &\nodata &\nodata &\nodata            &\nodata &\nodata   &\nodata                &0.5     &105\tablenotemark{a}  &29.2\\ 
NGC 5371                  &\nodata &\nodata  &\nodata     &\nodata &\nodata &\nodata                  &0.5     &217                   &6.2  &9.0     &137                   &0.9\\ 
                          &\nodata &\nodata  &\nodata     &\nodata &\nodata &\nodata                  &0.5     &107\tablenotemark{a}  &15.1 &0.5     &197                   &0.8\\ 
NGC 5806                  &\nodata &\nodata  &\nodata     &\nodata &\nodata &\nodata                  &14.0    &96\tablenotemark{a}   &1.6  &20.0    &96\tablenotemark{a}   &3.7\\
                          &\nodata &\nodata  &\nodata     &\nodata &\nodata &\nodata                  &1.0     &101                   &0.7  &6.0     &96\tablenotemark{a}   &1.2\\
NGC 6300                  &\nodata &\nodata  &\nodata     &\nodata &\nodata &\nodata                  &5.0     &174                   &9.5  &20.0    &109                   &9.3\\ 
                          &\nodata &\nodata  &\nodata     &\nodata &\nodata &\nodata                  &0.5     &209\tablenotemark{a}  &7.0  &8.0     &119\tablenotemark{a}  &4.6\\ 
NGC 7083                  &\nodata &\nodata  &\nodata     &\nodata &\nodata &\nodata            &\nodata &\nodata  &\nodata                 &13.0    &126                   &1.1\\  
                          &\nodata &\nodata  &\nodata     &\nodata &\nodata &\nodata            &\nodata &\nodata  &\nodata                 &0.5     &361                   &0.9\\ 
NGC 7217                  &\nodata &\nodata  &\nodata     &\nodata &\nodata &\nodata            &\nodata &\nodata  &\nodata                 &20.0    &176                   &15.6\\ 
                          &\nodata &\nodata  &\nodata     &\nodata &\nodata &\nodata            &\nodata &\nodata  &\nodata                 &19.0    &106\tablenotemark{a}  &4.0\\
NGC 7606                  &17.5    &142\tablenotemark{a} &1.7   &9.5     &142\tablenotemark{a}  &2.5    &20.0    &157                  &1.4 &20.0    &192                   &2.0\\
                          &0.5     &447                  &3.0   &0.5     &142\tablenotemark{a}  &3.6    &2.5     &357                  &2.7 &12.0    &187                   &2.3\\
\enddata
\tablenotetext{a}{\,Limit of parameters searched.}
\tablecomments{The first and second rows for each galaxy list information
for the uncontracted and contracted halo models.}
\end{deluxetable}

\begin{deluxetable}{cccccccccc}
\tabletypesize{\small}
\tablehead{
\colhead{} &
\colhead{} &
\colhead{} &
\colhead{} &
\colhead{} &
\multicolumn{5}{c}{Correlation Coefficients}\\
\cline{6-10}
\colhead{} &
\colhead{} &
\colhead{} &
\colhead{} &
\colhead{Galaxy Property} &
\multicolumn{2}{c}{All Data} &
\colhead{} &
\multicolumn{2}{c}{Without Outliers\tablenotemark{a}}\\
\cline{6-7}
\cline{9-10}
\colhead{} &
\colhead{} &
\colhead{} &
\colhead{} &
\colhead{} &
\colhead{$R_X/h_{IR}$} &
\colhead{$R_X$} &
\colhead{} &
\colhead{$R_X/h_{IR}$} &
\colhead{$R_X$} 
}
\startdata
&&&&$V_{tot,max}$ &0.41     &\nodata &  &0.90    &\nodata\\
&&&&$V_{b,max}$   &0.85     &\nodata &  &0.96    &\nodata\\
&&&&$R(V_{b,max})$   &\nodata     &0.22 &  &\nodata    &0.58\tablenotemark{b}\\
&&&&T-type        &0.89     &\nodata &  &0.90    &\nodata\\
&&&&log$_{10}M_b$ &0.65     &\nodata &  &0.90    &\nodata\\
&&&&$h_{IR}$      &\nodata  &0.21    &  &\nodata &0.80\tablenotemark{c}\\
&&&&$R_{10}$      &\nodata  &0.94    &  &\nodata &0.98\\
&&&&$B$           &0.36     &\nodata &  &0.77    &\nodata\\
&&&&$K$\tablenotemark{d}           &0.41     &\nodata &  &0.80    &\nodata\\
&&&&$\mu_{o,B}$   &0.51     &\nodata &  &0.68    &\nodata\\
&&&&$\mu_{o,K}$   &0.64     &\nodata &  &0.87    &\nodata\\
\enddata
\tablenotetext{a}{Outliers are NGC\,2841 and NGC\,4138.}
\tablenotetext{b}{Outliers for this relation are NGC\,3319 and
NGC\,4062.}
\tablenotetext{c}{NGC\,3319 is not included because it is an outlier
in this relation.}
\tablenotetext{d}{NGC\,3319 is not included because it does
not have $K$-band surface brightness
profiles with high enough signal-to-noise to derive an integrated magnitude.}
\end{deluxetable}

\end{document}